\documentclass[12pt]{article}
\pdfoutput=1
\usepackage{jheppub}
\usepackage{slashed}
\usepackage{amsmath}
\usepackage{amsfonts}
\usepackage{amssymb}
\usepackage{graphicx}
\usepackage[export]{adjustbox}
\newcommand{\bmat}{\left(\begin{array}}
\newcommand{\emat}{\end{array}\right)}

\def\yzero{\smash{\hbox{$y\kern-4pt\raise1pt\hbox{${}^\circ$}$}}}

\def\beq{\begin{equation}}
\def\eeq{\end{equation}}
\def\beqa{\begin{eqnarray}}
\def\eeqa{\end{eqnarray}}

\def\-{\hphantom{-}}
\def\ov{\overline}
\def\s2{\frac{1}{\sqrt2}}

\def\beq{\begin{equation}}
\def\eeq{\end{equation}}
\def\beqa{\begin{eqnarray}}
\def\eeqa{\end{eqnarray}}

\def\II{\relax{\rm I\kern-.18em I}}

\def\Dsl{\,\raise.15ex\hbox{/}\mkern-13.5mu D} %this one can be subscripted
\def\aD9{{\ov{\rm D9}}}

\def\IZ{{\bf {Z}}}

\def\NN{{\cal {N}}}

\def\Z{{\bf {Z}}}

\newcommand{\drawsquare}[2]{\hbox{%
\rule{#2pt}{#1pt}\hskip-#2pt%  left vertical
\rule{#1pt}{#2pt}\hskip-#1pt%  lower horizontal
\rule[#1pt]{#1pt}{#2pt}}\rule[#1pt]{#2pt}{#2pt}\hskip-#2pt%  upper horizontal
\rule{#2pt}{#1pt}}% right vertical

\newcommand{\fund}{\,\raisebox{-.5pt}{\drawsquare{6.5}{0.4}}\,}%  fund
\newcommand{\asymm}{\,\raisebox{-3.5pt}{\drawsquare{6.5}{0.4}}\hskip-6.9pt%
        \raisebox{3pt}{\drawsquare{6.5}{0.4}}\,}%  antisymmetric second rank

\newcommand{\antiasymm}{\overline{\asymm}}

\catcode`\@=11   
\newdimen\@rotdimen
\newbox\@rotbox  

\def\@vspec#1{\special{ps:#1}}%  passes #1 verbatim to the output
\def\@rotstart#1{\@vspec{gsave currentpoint currentpoint translate
   #1 neg exch neg exch translate}}% #1 can be any origin-fixing transformation
\def\@rotfinish{\@vspec{currentpoint grestore moveto}}% gets back in synch 
\def\@rotr#1{\@rotdimen=\ht#1\advance\@rotdimen by\dp#1%
   \hbox to\@rotdimen{\hskip\ht#1\vbox to\wd#1{\@rotstart{90 rotate}%
   \box#1\vss}\hss}\@rotfinish}
\def\@rotl#1{\@rotdimen=\ht#1\advance\@rotdimen by\dp#1%
   \hbox to\@rotdimen{\vbox to\wd#1{\vskip\wd#1\@rotstart{270 rotate}%
   \box#1\vss}\hss}\@rotfinish}%
\def\@rotu#1{\@rotdimen=\ht#1\advance\@rotdimen by\dp#1%
   \hbox to\wd#1{\hskip\wd#1\vbox to\@rotdimen{\vskip\@rotdimen
   \@rotstart{-1 dup scale}\box#1\vss}\hss}\@rotfinish}%
\def\@rotf#1{\hbox to\wd#1{\hskip\wd#1\@rotstart{-1 1 scale}%
   \box#1\hss}\@rotfinish}%
\def\rotate{\@ifnextchar[{\@rotate}{\@rotate[l]}}
\def\@rotate[#1]#2{\setbox\@rotbox=\hbox{#2}\@nameuse{@rot#1}\@rotbox}

\catcode`\@=12
\begin{document}

%----------------------------------------------------------------------%
%  numbering equations with section number
%----------------------------------------------------------------------%
\makeatletter
\@addtoreset{equation}{section}
\makeatother
\renewcommand{\theequation}{\thesection.\arabic{equation}}
%----------------------------------------------------------------------%
%  title page
%----------------------------------------------------------------------%
\pagestyle{empty}
%\vspace*{0.5in}
\rightline{IFT-UAM/CSIC-26-35 }
\rightline{ZMP-HH/26-7}
\vspace{.5cm}
\begin{center}
\Large{\bf The Art of Branching:\\ 
Cobordism Junctions of 10d String Theories
}
\\%[8mm] 
%}\\

\large{Chiara Altavista${}^1$, Edoardo Anastasi${}^1$, \\Roberta Angius${}^2$, Angel M. Uranga${}^1$ \\[4mm]}
\footnotesize{${}^1$ Instituto de F\'{\i}sica Te\'orica IFT-UAM/CSIC,\\[-0.3em] 
C/ Nicol\'as Cabrera 13-15, 
Campus de Cantoblanco, 28049 Madrid, Spain}\\ 
\footnotesize{${}^2$ II. Institut f\"ur Theoretische Physik, Universit\"at Hamburg, \\ Notkestrasse 9, 22607 Hamburg, Germany}\\ 
\footnotesize{\href{mailto:chiara.altavista@estudiante.uam.es}{chiara.altavista@estudiante.uam.es}}, \href{mailto:edo.anastasi@virgilio.it}{edo.anastasi@virgilio.it}, \\\href{mailto:roberta.angius@uni-hamburg.de}{roberta.angius@uni-hamburg.de}, \href{mailto:angel.uranga@csic.es}{angel.uranga@csic.es}

\vspace*{8mm}

\small{\bf Abstract} \\%[5mm]
\end{center}
\begin{center}
\begin{minipage}[h]{\textwidth}
\small{We describe the explicit construction of configurations of several 10d string theories joining at a 9d junction, providing a dynamical realization of cobordisms between multiple 10d string theories, predicted by the Cobordism Conjecture. We provide the microscopic worldsheet description of the configuration in a generalization of the `going up and down the RG flow' interpolations recently used in the description of IIA/IIB domain wall. The interpolations involve additional degrees of freedom, which are gapped except at the branch point, at which the gap closes and triggers the branching transition. The extra degrees of freedom admit an interpretation in terms of additional dimensions in a supercritical string theory, which reduces to the 10d junction configuration upon closed tachyon condensation. Quantum corrections of the 2d worldsheet theory turn the junction into a strongly coupled lightlike core whose UV resolution lies beyond worldsheet techniques. We construct explicit examples of junctions of 10d heterotic string theories, type 0, and type II theories and orientifolds thereof. Our explicit examples include junctions of 10d chiral theories whose chiral fields flow between different branches. One particularly nice configuration is a 4-branch junction of the IIB theory, with type I, the non-supersymmetric $USp(32)$ theory and the $U(32)$ orientifold of 0B theory, thus assembling the four non-tachyonic descendants of type 0B theory. 
}

\newpage

\end{minipage}
\end{center}
\newpage
%----------------------------------------------------------------------%
%  Resetting of counters
%----------------------------------------------------------------------%
\setcounter{page}{1}
\pagestyle{plain}
\renewcommand{\thefootnote}{\arabic{footnote}}
\setcounter{footnote}{0}
%----------------------------------------------------------------------%
%  Paper begins
%----------------------------------------------------------------------%

\vspace*{-1cm}

\tableofcontents

\vspace*{1cm}

%\newpage

\section{Introduction}
\label{sec:intro}

One of the central ideas driving progress in string theory (both within the swampland program \cite{Vafa:2005ui} and more broadly) is the cobordism conjecture \cite{McNamara:2019rup}. It posits that the (suitably defined) cobordism charges of any consistent quantum gravity configuration should be trivial. This implies the existence of an extremely rich set of spacetime dynamical configurations connecting different theories, specially interesting and non-trivial when applied to string theories or M-theory in their maximal spacetime dimension. One instance is given by end of the world (ETW) boundaries at which spacetime terminates, like the Horava-Witten ETW wall in 11d M-theory \cite{Horava:1995qa,Horava:1996ma} (see also \cite{Montero:2025ayi}), or the type I' O8-planes (plus D8-branes) \cite{Polchinski:1995df} as ETW boundaries of (possibly massive) 10d type IIA theory \footnote{See \cite{Buratti:2021yia,Buratti:2021fiv,Angius:2022aeq,Blumenhagen:2022mqw,Angius:2022mgh,Angius:2023xtu,Blumenhagen:2023abk,Angius:2023uqk,Angius:2024zjv,Delgado:2023uqk, Antonelli:2019aib}  for other dynamical realizations of cobordism ETW boundaries in diverse contexts.}. Another interesting class of configurations are 9d interfaces interpolating between 10d string theories, like the 10d type IIA/IIB domain wall (and its 10d 0A/0B cousin) whose properties and construction have been recently explored in \cite{Heckman:2025wqd,Anastasi:2026cus,Torres:2026vxx}. 

\begin{figure}[htb]
\begin{center}
\includegraphics[scale=.35]{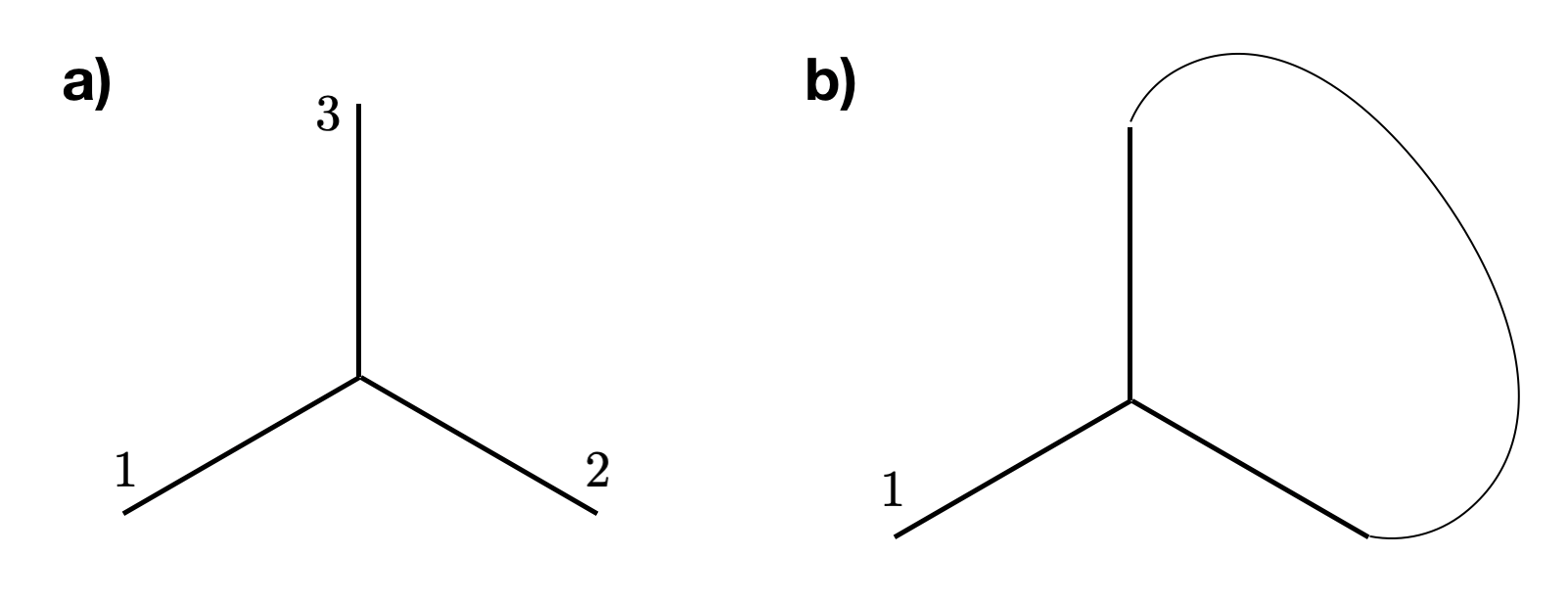}
\caption{\small a) Sketch of a 3-branch junction of 10d string theories. b) Depiction of the interplay between multi-branch junctions and ETW boundaries for a single branch.}
\label{fig:junction-general}
\end{center}
\end{figure}

There is however yet another kind of configurations which have not been considered in the literature so far (see however the early work \cite{Hellerman:2010dv}): configurations of more than two 10d string theories joining at a 9d junction, see Figure \ref{fig:junction-general}a. The aim of this paper is to provide explicit constructions of such junctions for various sets of 10d string theories. At the risk of giving away the punchline, we advance that we construct explicit junctions of (up to six) 10d type 0A or 0B theories; junctions of various non-supersymmetric heterotic theories, including a junction of three 10d non-supersymmetric (but chiral!) $E_8\times SO(16)$ heterotic theories; junctions of four type IIA or IIB theories; and a beautiful 4-branch junction of one 10d type IIB theory, one 10d type I theory, one 10d non-supersymmetric non-tachyonic $USp(32)$ theory, and one 10d non-tachyonic $U(32)$ orientifold of 0B theory, precisely the four 10d non-tachyonic theories related to type 0B theory.

%%%%%%%%%%%%
%%%%%%%%%%%%

The main reason why multi-branch junction configurations have not been considered in the literature is that they can be regarded as a cobordism between one 10d string theory in one branch and the disconnected sum of the 10d string theories in the remaining branches. It is non-trivial to make sense of this disconnected sum, and it is even less clear what tools one could use to even formulate them consistently. We advocate that such junction configurations can be appropriately defined in terms of a worldsheet formulation, which encompasses all the string theories in the junction. 

The worldsheet approach is analogous to that successfully employed in  \cite{Anastasi:2026cus} to describe the 10d type IIA/IIB domain wall (and its 0A/0B cousin). In the context of 9d interfaces connecting 10d string theories, the key concept is regarding the 2d worldsheet CFTs of the theories as the IR limit of non-trivial RG flows from UV 2d QFTs, which are connected via an interpolating parameter deforming the UV lagrangian. The worldsheet version of the cobordism thus corresponds to a transition along a one-parameter trajectory moving up and down the RG flow in the sense of \cite{Gaiotto:2019asa} (analogous to the application in the heterotic context with the notion of cobordism charges formalized in terms of elliptic genera or topological modular form classes \cite{Witten:1986bf, Gukov:2018iiq,Gaiotto:2019asa,Tachikawa:2021mvw,Tachikawa:2021mby,Tachikawa:2023nne,Tachikawa:2025awi,Moore:2025tmt}). Actually, the dynamical realization in \cite{Anastasi:2026cus} motivated enlarging the class of interpolations by allowing them to cross non-compact CFTs, thus permitting non-trivial changes in the topological order of the 2d theory. The dynamical realization of these interpolations in \cite{Anastasi:2026cus} simply corresponds to letting the parameter depend on a 10d spacetime coordinate, $\lambda(X)$,  and explicit examples of such dynamical realizations were provided, in which the additional degrees of freedom admit an interpretation in terms of supercritical strings.

Our worldsheet construction of junctions shares a similar spirit, with one extra ingredient, namely {\em branching}. A simplified description of how this works is as follows. We start with one 2d CFT$_1$ describing a 10d string theory, and move up the RG flow to a UV theory QFT$_1$ with a parameter $\lambda_1$. As we move this parameter in some coupling space, we hit a branching point, at which two extra parameters $\lambda_2$, $\lambda_3$ appear, which are mutually incompatible (among themselves and with $\lambda_1$): turning on one of them makes the others disappear. We may hence continue the interpolation by turning on $\lambda_2$, with $\lambda_3$ (and $\lambda_1$) disappearing, or viceversa. By letting the corresponding theories QFT$_2$ and QFT$_3$ run down the RG flow, we end up with two theories CFT$_1$, CFT$_2$, describing two 10d disconnected string theories. This is the worldsheet version of a spacetime junction. By considering a 3-branch junction of spacetime and letting the parameters depend on the corresponding spacetime coordinates, we have a dynamical realization of the junction. 

Our actual realizations include an extra subtlety: given that in the physical realization the parameters depend on the coordinates themselves, they effectively turn into 2d fields, so that the mutual incompatibility of the two parameters/fields $\lambda_2\sim X$ and $\lambda_3\sim Y$ translates into the fact that a vev for one makes the other massive, and viceversa. A simple realization, arising as a subsector in our configurations, is the following: consider (a sector) of our CFT$_1$ to be a 2d $(1,1)$ theory with one free real superfield $Z$. For generic $Z$, this can be seen as the IR limit of a theory with two extra real superfields $X$, $Y$ with superpotential $W=ZXY$. However, at $Z=0$ the gap for $X$, $Y$ goes to zero, and we can move in those directions. Concretely, we may move into the $X$-branch, by giving a vev to $X$, which makes $Y$ and $Z$ massive, or move into the $Y$-branch, by giving a vev to $Y$, which makes $X$ and $Z$ massive. The resulting theories in the IR limit in the $X$- and $Y$-branches describe two disconnected CFTs. 

This simple system is the backbone of our construction of junction configurations of 10d string theories. In fact, in order to integrate it in a well defined physical realization, our constructions are based on supercritical string theories\footnote{We note that, although the no-ghost theorem has not been proven for supercritical strings, there is no obvious obstruction to doing so; see \cite{Bautista:2019jau} for related recent advances.} and closed tachyon condensation (see \cite{Hellerman:2006nx,Hellerman:2006hf,Hellerman:2006ff,Hellerman:2007fc,Hellerman:2007ym,Hellerman:2007zz,Hellerman:2008wp,Hellerman:2010dv,Berasaluce-Gonzalez:2013sna,Garcia-Etxebarria:2014txa,Garcia-Etxebarria:2015ota,Angius:2022mgh} for related work). This simply amounts to a reinterpretation of the massive degrees of freedom as parametrizing extra spacetime coordinates, so it is not an absolutely essential ingredient in the construction. On the other hand, it is a fairly convenient one, as it facilitates the interpretation of certain more involved operations (like orbifold and orientifold quotients) on these extra degrees of freedom.
In any event, all our constructions can be undressed of the supercritical envelope and understood as the inclusion of extra gapped sectors, which allow for interpolations (by going up and down the RG flow) of CFTs passing through branch points.

Let us note that, in analogy with \cite{Anastasi:2026cus}, at the branching point the interpolation crosses a non-compact CFT. This feature allows for changes in the topological structure of the theories, and allow the junctions to connect different 10d string theories. Relatedly, and also as in \cite{Anastasi:2026cus}, the quantum effects upon integrating out massive fields renormalize the spacetime metric and dilaton background, and lead to a singularity at the branching point. This drives the junction locus to infinite spacetime distance in the string frame, and turns it into a strongly coupled lightlike core. The ultimate fate of the junction thus lies beyond the reach of our worldsheet techniques, and depends on the possible UV resolution of the strong coupling singularity. This may end up leading to totally disconnected branches (so that the former junction effectively ends up as a set of lightlike ETW walls for the theories on the branes), or may allow from some transmissible junction. We remain agnostic regarding the possible UV resolution of the strongly coupled core, but interestingly find that several junctions seem ready to admit the smooth propagation of 10d chiral fields (which are topologically protected against disappearing) allowing to define concrete junction conditions for these fields.

The presence of chiral fields in some of our junctions is important from yet a different, but related, perspective. Given e.g. a 4-branch junction where the different branches join like the four half-lines in the equation $XY=0$ (i.e. the $X$- and $Y$-axis), it is geometrically possible to consider perturbations deforming it to $XY=\epsilon$, which describes two disconnected spacetimes. This can be prevented by quotienting by additional $\IZ_2$ symmetries (such as $X\to -X$, or $Y\to -Y$, or both), under which the deformation is not invariant. Such quotients are often present in the junctions we construct, and the robustness of the junction manifests in the appearance of chiral fields supported on the 10d spacetimes, i.e. the appearance of chiral 10d string theories on some of the branches, as we now explain.

In general, the interplay between cobordism and chirality is a fascinating arena. For instance, the existence of ETW branes for 10d chiral string theories implies that there must exist symmetric mass generations mechanisms, physical processes which can gap the whole set of chiral anomaly-free fields, even if chirality forbids mass terms (see \cite{Angius:2024pqk} for the construction of boundary configurations for 6d and 4d chiral gravitational theories, and e.g. \cite{Razamat:2020kyf,Tong:2021phe,Wang:2022ucy} for 2d and 4d QFT realizations of symmetric mass generation). Such mechanisms operate at strong coupling, and should underlie the nature of the (still unknown) ETW boundaries for the 10d chiral string theories, including type IIB, type I, the heterotics, etc. Similarly, in 9d interfaces between 10d string theories differing in their chiral spectrum (like the IIA/IIB domain wall \cite{Heckman:2025wqd,Anastasi:2026cus,Torres:2026vxx}), the wall acts as a boundary for each individual side, demanding again symmetric mass generation mechanisms. In contrast, in multi-branch junctions of different 10d string theories, which in general can differ in their chiral spectrum, there is a novel possibility: rather than being gapped, the chiral content of the 10d string theory in any branch may continue existing across the junction by having different chiral fields propagate into several other branches. We refer to this mechanism as the chiral flow of the junction. Junctions with a non-trivial chiral flow are robust against deformation of the kind discussed in the previous section, so they naturally arise in the presence of addition orbifold and/or orientifold quotients. To highlight the special status and beauty of configurations with non-trivial chiral flows, we refer to them a {\em bouquets}, as they assemble a non-trivial set of {\em flowers} (i.e. chiral fields that flow across different theories in the junction). We will find explicit examples of chiral bouquets, including a junction of 10d type IIB, type I, the $USp(32)$ theory, and the $U(32)$ orientifold of 0B.

Let us note that general junctions of 10d string theories may be relevant to discuss even richer networks of 10d string theories, for instance by including several junctions glued along one or several branches. Note also that this gluing procedure can be used to reduce the number of branches, by gluing several branches in a single junction. This can be relevant to relate our constructions with more familiar cobordism configurations, such as ETW boundaries or interfaces. For instance, if the 10d theories in two branches in a junction are the same (up to orientation flip, trivial for non-chiral theories, but crucial for chiral ones), it should be topologically allowed to glue them into a closed loop and obtain a configuration with less branches, or even with a single one, effectively describing an ETW boundary configuration for the left-over theory, see Figure \ref{fig:junction-general}b. Therefore, the junctions we construct may yield novel implications regarding the construction of still unknown ETW boundaries or interfaces for 10d string theories. Even though this should be topologically possible, this work focuses on constructing explicit dynamical junction configurations, and we emphasize as a disclaimer that we have not found an explicit example including closed loops.

The paper is organized as follows. In Section \ref{sec:type0-typeii} we provide a prototype of our construction technique for junctions of 10d string theories, illustrating it with junctions for 10d type 0 theories (section \ref{sec:type0-junctions}) --with a discussion of the effects of quantum corrections (section \ref{sec:quantum})--, and 10d type II theories (section \ref{sec:typeII-junctions}). In Section \ref{sec:heterotic-junctions} we build explicit junctions of 10d heterotic string theories, and study several classes of such constructions: in section section \ref{sec:basic-heterotic-junctions} we build heterotic junctions using only fermionic $\IZ_2$ gaugings on the worldsheet; section \ref{sec:4-branch-junction-heterotic} uses one $\IZ_2$ orbifold action on the extra (supercritical) bosonic degrees of freedom; in section \ref{sec:3-branch-junction-heterotic} we use a $\IZ_2\times\IZ_2$ orbifold acting on the extra bosonic degrees of freedom and build a 3-branch junction of three 10d non-supersymmetric (but chiral) $E_8\times SO(16)$ theories, exhibiting a non-trivial chiral flow, i.e. a heterotic bouquet configuration. In Section \ref{sec:bouquet} we construct an even more remarkable bouquet configuration, as a 4-branch junction of type IIB, type I, the $USp(32)$ theory, and the $U(32)$ orientifold of type 0B. In section \ref{sec:orientifold-group} we describe the orbifold and orientifold actions involved in the construction, in section \ref{sec:orientifold-theories} we derive the nature of the resulting 10d string theories on the different branches, and in section \ref{sec:bouquet-flow} we describe the non-trivial chiral flow in the bouquet. In Section \ref{sec:conclusions} we offer some final remarks. To make this works self-contained, we provide several appendices with additional details. In Appendix \ref{app:critical} we quickly review the construction of 10d heterotic string theories, focusing on the less familiar non-supersymmetric ones. Appendix \ref{app:supercritical} reviews supercritical type 0 theories and their closed tachyon condensation to 10d type 0 and type II theories. Appendix \ref{app:supercritical-heterotic} reviews supercritical heterotic theories (including known, as well as some new, models) and their closed tachyon condensation to 10d (supersymmetric or not) heterotic theories. Appendix \ref{app:quantum} provides some details on the computation to the quantum corrections of the 2d theories. Finally, Appendix \ref{app:correlators} gives extra details on the computation of the junction conditions from 2d 3-point correlators for the heterotic bouquet configuration.

\section{Junctions of 10d type 0 or type II theories}
\label{sec:type0-typeii}

In this section we provide the basic concept of how to define junctions of 10d string theories, in terms of type 0 theories, or of type II theories. Generalizations to heterotic theories will be described in sections \ref{sec:heterotic-junctions}, and to unoriented theories in \ref{sec:bouquet}. As explained in the introduction, we describe the worldsheet theory of the configuration in terms of supercritical string and closed tachyon condensation, but the supercritical sector of the CFT may be also regarded as a simple addition of (generically) gapped degrees of freedom as we go up and down the RG flow in the 2d worldsheet theory.

\subsection{Junctions of 10d type 0 theories}
\label{sec:type0-junctions}

As explained, we carry out our construction in terms of supercritical strings, see appendix \ref{app:supercritical} for type 0 supercritical strings in $D=10+n$ dimensions. 
Let us start with 12d supercritical type 0 theory, whose worldsheet theory (in the light-cone gauge) is given by a $(1,1)$ supersymmetric theory with 10 multiplets. Since we are constructing a 9d interface in spacetime, we split them as a set of 9 bosons $X^i$, and 3 additional ones denoted by $X$, $Y$, $Z$, and similarly for the fermions $\psi^i,{\tilde\psi}^i$, $\psi_X,{\tilde \psi}_X$, $\psi_Y,{\tilde \psi}_Y$, $\psi_Z,{\tilde \psi}_Z$, where untilded/tilded fields correspond to the left-/right-moving sectors. We introduce a superpotential (i.e. a tachyon profile) with the structure
\beqa
\Delta{\cal L}=\int d^2\theta\, XYZ\, .
\label{supo-xyz}
\eeqa
By abuse of language, we use the same letters to denote the full $(1,1)$ multiplets and their bosonic components. In the above formula we have ignored for simplicity the exponential in the lightlike coordinate $X^+$ (introduced in \cite{Hellerman:2006ff} to make the above deformation a marginal operator, see Appendix \ref{app:supercritical} for detail). If included, the following statements should be regarded as holding at late $X^+$.

\begin{figure}[htb]
\begin{center}
\includegraphics[scale=.35]{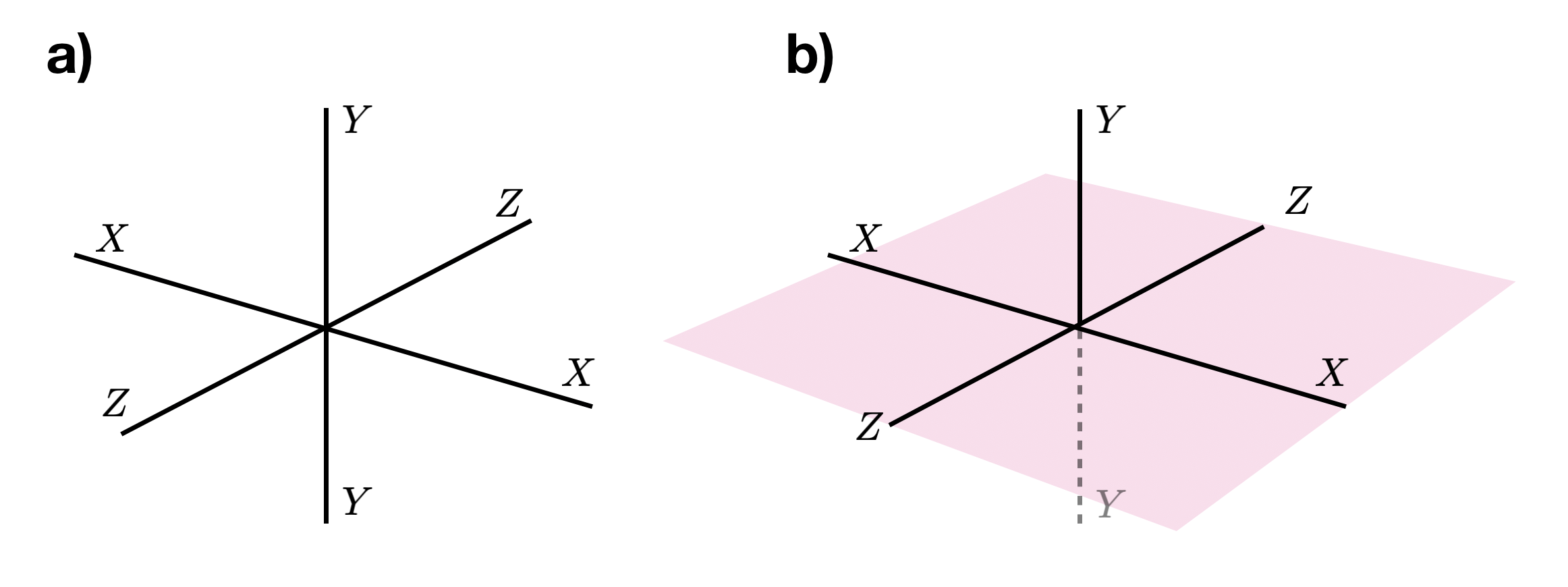}
\caption{\small a) Structure of the junction of 10d type 0 theories resulting from closed tachyon condensation. Some of the branches may be removed by higher order deformations of the superpotential.
b) Structure of the junction of 10d type II and type 0 theories after an orbifold, whose fixed plane is shown in pink. 
The dashed line correspond to the $Y<0$ image of the $Y>0$ branch under it. The $Y$-branch may be removed by higher order deformations of the superpotential.}
\label{fig:junction-type0-ii}
\end{center}
\end{figure}

The resulting terms in components, c.f. (\ref{wsupo}), (\ref{pot}) include a scalar potential and fermion Yukawa interactions with the structure 
\beqa
 V(X,Y,Z)&\, \sim\, &  X^2Y^2+Y^2Z^2+Z^2X^2\nonumber \\
{\cal L}_{\rm ferm}  &\,\sim\, & X (\psi_Y{\tilde \psi}_Z+\psi_Z{\tilde \psi}_Y) + Y (\psi_Z{\tilde \psi}_X+\psi_X{\tilde \psi}_Z)+Z (\psi_X{\tilde \psi}_Y+\psi_Y{\tilde \psi}_X)\, .\quad
\label{couplings-xyz-type0}
\eeqa
The classical moduli space\footnote{We will eventually use this as the moduli space of the quantum theory, in the usual sense of the Born-Oppenheimer approximation.} is given by three branches, dubbed the $X$-branch (for which $Y=Z=0$, $X\neq 0$), the $Y$-branch ($X=Z=0$, $Y\neq 0$), and the $Z$-branch ($X=Y=0$, $Z\neq 0$). Along e.g. the $Z$-branch, the $(1,1)$ $X$ and $Y$ multiplets (including bosons and fermions) become massive. In the closed tachyon condensation interpretation we fall into a 10d critical type 0 theory with 10d coordinates $X^i$, $Z$ (and $X$, $Y$ are interpreted as supercritical extra dimensions, removed by the closed tachyon profile). Clearly, a similar pattern holds in the $Y$- and $Z$-branches, with the corresponding unconstrained coordinate becoming the 10$^{th}$ spacetime components. Amusingly, dimensions which are part of the physical 10d spacetime in one branch correspond to gapped supercritical dimensions in the othes. The junction configuration is shown in Figure \ref{fig:junction-type0-ii}a.

%%%%%%%%%%%%
%%%%%%%%%%%%

In the spirit of \cite{Anastasi:2026cus}, the above superpotential may be regarded as the local approximation to a more involved functional dependence. In particular it is easy to modify it so that along e.g. the $Z$-branch the tachyon profile tends to a constant, by symply replacing $Z\to \arctan Z$ in the above formula (and analogously for the other coordinates, even simultaneously). In such situation, the tachyon condensation  at large $Z$ along the $Z$-branch effectively becomes identical to that of constant tachyon profiles in \cite{Hellerman:2006ff}, c.f. appendix \ref{app:type0}.

We note that the above 2d theory is subject to quantum corrections. However, it is easy to see that, along each branch, the fields becoming massive complete full $(1,1)$ multiplets, and the fermion-boson degeneracy implies the absence of corrections to the scalar potential. Therefore the only quantum corrections modify at most the kinetic terms, redefining the spacetime metric (and, as we will see, dilaton) background. This effect will be discussed in section \ref{sec:quantum}.

In this construction, the theories living at the different branches are all 10d type 0 theories. We will later see that it is possible to have different 10d string theories in different asymptotic branches away from the 9d junction. We note that it is in principle even possible to have two different theories in the two regions $Z>0$, $Z<0$ of the $Z$-branch. For this reason, we prefer to regard the above configuration as a 6-branch junction of 6 type 0 theories, each defined on a `half' 10d spacetime ending on the 9d junction. 

We also note that it is possible to deform the superpotential (\ref{supo-xyz}) to remove some of the asymptotic branches. Consider the modification to
\beqa
W=XYZ+Y^3\, ,
\label{y-cube-deformation}
\eeqa
(we consider deformations odd in $Y$ for future applications to junctions of type II theories, where the superpotential must be odd in the corresponding coordinate). Then it is easy to see that the $X$- and $Z$-branches are unchanged, but the $Y$-branch has been lifted. We therefore end up with a 4-branch junction of 4 10d type 0 theories.

Let us finally note that another possible way to reduce the number of branches in the junction is to orbifold the configurations by some geometric action, for instance flipping the sign of some of the coordinates $X,Y,Z$. In fact, such orbifolds are natural from the viewpoint of supercritical extensions of the 10d supersymmetric theories, see appendices \ref{app:typeII}, \ref{sec:supercritical-heterotic-orbifold-yes}. We will come back to this possibility in explicit constructions in sections \ref{sec:typeII-junctions}, \ref{sec:heterotic-junctions} and \ref{sec:bouquet}.

\subsection{Quantum corrections}
\label{sec:quantum}

As anticipated in the previous section, the above theory has quantum corrections. These can be computed e.g. along the $Z$-branch, by integrating out the massive multiplets $X,Y$. Due to supersymmetry, there is no correction to the scalar potential, and the only non-trivial corrections modify the kinetic term for $Z$, i.e. $g_{zz}(\partial Z)^2$. This is not protected against supersymmetry, and by dimensional analysis the correction must be of the form
\beqa
g_{ZZ}\sim \frac 1{Z^2}\, .
\eeqa
The computation is shown in appendix \ref{app:quantum}. Interpreting this as a correction to the spacetime metric, this implies that the position of the 9d branch point $Z=0$ is pushed to infinite distance in the string frame metric. A more precise picture is obtained by dressing the superpotential (\ref{supo-xyz}) with a lightlike exponential in $X^+$, as in \cite{Hellerman:2006ff} (see appendix \ref{app:supercritical})
\beqa
W=\mu e^{\beta X^+} Z(X'{}^2-Y'{}^2)\, ,
\eeqa
where $\mu$ is an arbitrary parameter, $\beta$ is fixed to make the coupling marginal, and we have redefined $X'=(X+Y)/2$, $Y'=(X-Y)/2$. 
In the $Z$-branch, regarding the value of $Z$ as a fixed background, the system is therefore identical to those in \cite{Hellerman:2006ff} for the removal of two spacetime coordinates. As explained there, the 1-loop computation is exact, and yields a correction to the spacetime metric and dilaton background given (for each coordinate integrated out) by
\begin{equation}
\delta G_{\mu\nu} =  \frac{\alpha'}{4} \frac{\partial_\mu M \partial_\nu M}{M^2},\quad \Delta\Phi = \frac14\ln\left(\frac{M}{\tilde{\mu}}\right)\, ,
\end{equation}
where $\tilde{\mu}$ is a renormalization scale (setting the value of the dilaton at a reference point). In our case, the computation is similar (save for the doubling of massive fields) to that in \cite{Anastasi:2026cus}, which adapts that in \cite{Hellerman:2006ff} for a field-dependent mass. In our case the mass of $X'$ and $Y'$ is (up to sign)
\begin{equation}
M= 2Z\, e^{\beta X^+}\, .
\end{equation}
The renormalization of the full set of components of the metric and the dilaton are as in \cite{Hellerman:2006ff,Anastasi:2026cus}, namely
\begin{equation}  
\delta G_{++} =\frac{\alpha'}{4}\beta^2\quad ,\quad  \delta G_{ZZ} =\frac{\alpha'}{4}\frac{1}{Z^2},\quad \delta G_{9+} =\frac{\alpha'}{4}\frac{\beta}{Z},\quad \Delta\Phi= \frac{1}{4}\left(\beta\, X^+ + \ln(2Z)\right).
\end{equation}
The resulting background can be recast, by a change of coordinates explained in \cite{Anastasi:2026cus}, as a metric in which the naive junction point $Z=0$ has been sent to infinite (string frame) distance, and the junction core is replaced by a lightlike throat supported by a lightlike linear dilaton background.

The final result is thus a lightlike linear dilaton background in the 10d critical theory of the $Z$-branch. This is no surprise, since, following \cite{Anastasi:2026cus}, it is the only reasonable outcome of the integrating out process once we realize we need a 9d Poincar\'e invariant configuration in the 10d critical theory, so there is not enough left-over central charge to accommodate any non-trivial CFT.

In the Einstein frame, there is a singularity at strong coupling at finite distance (actually, two such back to back singularities for the $Z>0$ and $Z<0$ branches), whose resolution is beyond the reach of our worldsheet techniques. Clearly, a similar pattern holds for the $X$- and $Y$- branches, so the final configuration is given by 6 type 0 theories defined on 10d spaces bounded by 9d light-like strongly coupled singularities, at which they join. The nature of the joining, and in particular if the corresponding junction is traversable or not, depends on the non-perturbative completion of the configuration. In this work we will be agnostic about this completion, and simply work with the underlying theory before integrating out heavy modes, with the assumption that this will flow to a singular CFT at the 1d defect core which will require non-perturbative completion. Interestingly, we will find examples where the theories in the junction are `ready' for it to admit a transmissible UV resolution, in the sense explained in the introduction.

\subsection{Junctions of 10d type II theories}
\label{sec:typeII-junctions}

The configurations in section \ref{sec:type0-junctions} provide an interesting class of multi-branched junctions of  10d type 0 string theories.  
In this section we describe a simple modification by an orbifold, which provides the construction of junctions with 10d supersymmetric type IIA or type IIB theories in some of its branches. The extra orbifold action on the supercritical directions, turning some type 0 theories into type II ones, is reviewed in appendix \ref{app:typeII} (see \cite{Hellerman:2006ff} for details).

\subsubsection{Junctions of four 10d type II theories}
\label{sec:ii-ii}

We start with the $(1,1)$ worldsheet theory of section \ref{sec:type0-junctions} describing 12d supercritical type 0 theory, and split the corresponding 10 coordinates (in the light-cone gauge) as 9 $X^i$ (with left- and right-moving fermion partners $\psi^i$, ${\tilde \psi}^i$), plus 3 denoted by $X$, $Y$, $Z$ (with left-moving fermion partners $\psi_X$, $\psi_Y$, $\psi_Z$, and the tilded versions for the right-moving ones). We implement a closed tachyon condensation (\ref{supo-xyz}) that leads to a 6-branch junction of 10d type 0 theories. The theories are 0A or 0B depending on the starting 12d type 0A or 0B theory.

We now implement a $\IZ_2$ orbifold to relate some of these 10d theories to type II versions. As explained in appendix \ref{app:typeII}, the orbifold must act on the ambient 12d spacetime by flipping one of the supercritical coordinates, which, without loss of generality, we take to be $Y$, and to anticommute with the left-moving supercurrent, as befits an R-parity. Explicitly the orbifold generator $g_{II}$ acts as
\beqa
&g_{II}:& (X^i;X,Y,Z)\to (X^i;X,-Y,Z)\nonumber \\
%\quad , \quad
&&({\tilde\psi}^i;{\tilde \psi}_X,{\tilde \psi}_Y,{\tilde \psi}_Z)\to ({\tilde\psi}^i;{\tilde \psi}_X,-{\tilde \psi}_Y,{\tilde \psi}_Z) \nonumber \\
&&(\psi^i;\psi_X,\psi_Y,\psi_Z)\to (-\psi^i;-\psi_X,\psi_Y,-\psi_Z) \, .
\label{orbifold-typeii}
\eeqa
It is easy to check that this action compatible with the tachyon condensation, i.e. preserves the interactions (\ref{couplings-xyz-type0}), or equivalently it acts by flipping the superpotential (\ref{supo-xyz}), as befits an R-parity.

After the tachyon condensation, the configuration contains the $X$-, $Y$- and $Z$-branches. The $X$- and $Z$-branches sit within the 11d orbifold fixed locus $Y=0$, and hence experience the orbifold projection. They therefore support 10d type II theories (of the A/B kind corresponding to the underlying type 0 theory). On the other hand, the $Y$-branch has its $Y<0$ and $Y>0$ parts mapped to each other by the orbifold, so the quotient contains only one copy (e.g. $Y>0$), with no orbifold projection imposed. This $Y>0$ branch thus continues supporting a 10d type 0 theory. The result of the configuration is a 5-branch junction of four 10d type II theories and one 10d type 0 theory, defined on 10d spacetimes with a 9d boundary, at which they are joined.

As anticipated in section \ref{sec:type0-junctions}, it is easy to modify the tachyon superpotential by including higher orders to get rid of the $Y$-branch, for instance by considering (\ref{y-cube-deformation}), which we repeat for convenience:
\beqa
W=XYZ+Y^3\, .
\label{y-cube-deformation-bis}
\eeqa
This superpotential is odd under the orbifold action $g_{II}$, so that the latter remains a good R-parity. The only relevant effect of the $Y^3$ deformation is to lift the $Y$-branch, leading to a 4-brach junction of four 10d type II theories (all of the same A/B kind, fixed by the underlying type 0 theory). 

Note that it is not possible to add higher order terms given by monomials in $X$ or $Z$ alone, because they are not odd under $g_{II}$. A complementary spacetime explanation of this fact is that, if applied to a type IIB setup, the removal of a branch is similar to gapping the 10d spacetime field content of the theory, a process whose description is highly non-trivial (and presently unknown) because it is highly chiral (see \cite{Angius:2024pqk} for the construction of boundary configurations for 6d and 4d chiral gravitational theories, and e.g. \cite{Razamat:2020kyf,Tong:2021phe,Wang:2022ucy} for 2d and 4d QFT realizations of symmetric mass generation).

In the above configuration, the 10d type II theories are paired up across the junction, so the chirality flow is straightforward. As explained in the introduction, this is related to the fact that the above junctions can be smoothed out by adding a deformation $\Delta W=\epsilon Y$ to the superpotential, so that the former locus $XY=0$ of the $X$- and $Z$-branches is resolved into  two non-intersecting real curves $XY=\epsilon$. Therefore, even though the junction contains branches with 10d chiral string theory, it does not qualify as a bouquet with non-trivial chiral flow, of the kind explained in the introduction. In section \ref{sec:bouquet} we will include an extra orientifold and get a configuration which does lead to a bouquet with non-trivial chiral flow. This is correlated with the fact that the extra quotient forbids the above kind of deformation of the intersection.

\section{Junctions of 10d heterotic theories}
\label{sec:heterotic-junctions}

In this section we describe the construction of junctions of 10d heterotic string theories. The construction is based on the supercritical heterotic theories reviewed in Appendix \ref{app:supercritical-heterotic}, in particular some variants not considered previously in the literature, discussed in appendix \ref{sec:heterotic-orbifold-variants}. As some of the participating 10d heterotics are less familiar, Appendix \ref{app:critical} reviews general 10d supersymmetric and non-supersymmetric heterotic theories.

\subsection{The basic junction of 10d heterotic theories}
\label{sec:basic-heterotic-junctions}

In this section we consider the basic junction of six 10d heterotic string theories, the heterotic analogue of that in section \ref{sec:type0-junctions} for type 0 and type II theories. We also describe modifications by including extra $\IZ_2$ gaugings by purely fermionic actions. This corresponds to introducing various generalized GSO projections, allowing for the appearance of various 10d heterotic string theories.

\subsubsection{The heterotic junction from closed tachyon condensation}

Let us start with the 12d supercritical heterotic theory with the diagonal modular invariant, see appendix \ref{sec:supercritical-heterotic-tachyons}. We split the 10 (light-cone gauge) bosons and right-moving fermions in the $(0,1)$ multiplets in a set of 7, with component boson and fermion fields $X^i$ and ${\tilde \psi}^i$, plus 3 with components $X$, $Y$, $Z$ and ${\tilde\psi}_X$, ${\tilde\psi}_Y$, ${\tilde\psi}_Z$; we also split the 34 left-moving fermions into 31 denoted by $\lambda_a$, plus 3, denoted (with some hindsight) by $\lambda_X$, $\lambda_Y$, $\lambda_Z$.

As explained in Appendix \ref{sec:supercritical-heterotic-tachyons}, the NSNS tachyon couples to the worldsheet theory as a $(0,1)$ superpotential (\ref{supo-heterotic-tachyon}), producing the fermion Yukawa couplings and scalar potential (\ref{heterotic-tachyon-condensation}). In order to reproduce a junction configuration, we choose a $(0,1)$ superpotential 
\beqa
\int d\theta_+ \left(\, \lambda_X\, YZ\,+\, \lambda_Y\, XZ\,+\, \lambda_Z\, XY\, \right)\, .
\eeqa
Using (\ref{heterotic-tachyon-condensation}), we get a scalar potential and fermions couplings with the structure
\beqa
 V(X,Y,Z)&\,\sim\, &X^2 Y^2+Y^2 Z^2+Z^2 X^2\nonumber \\
 {\cal L}_{ferm}&\, \sim\,& X (\lambda_Y{\tilde \psi}_Z+\lambda_Z{\tilde \psi}_Y) + Y (\lambda_Z{\tilde \psi}_X+\lambda_X{\tilde \psi}_Z)+ Z (\lambda_X{\tilde \psi}_Y+\lambda_Y{\tilde \psi}_X)\, .\quad
\label{xyz-couplings-heterotic}
\eeqa
These are exactly like in the type 0 case (\ref{couplings-xyz-type0}), with $\lambda_{X}$ playing the role of $\psi_X$, and similarly for the $Y$ and $Z$ subindices (hence the choice of subindices for the $\lambda$'s). Therefore this sector enjoys an accidental $(1,1)$ supersymmetry, which allows to borrow some of the type 0 results.

The scalar potential again leads to three branches: the $X$-branch (with $Y=Z=0$, $X\neq 0$), the $Y$-branch ($X=Z=0$, $Y\neq 0$), and the $Z$-branch ($X=Y=0$, $Z\neq 0$). Along e.g. the $Z$-branch, the $(0,1)$ multiplets $X$ and $Y$ become massive with the left-moving Fermi multiplets including $\lambda_X$ and $\lambda_Y$, whereas $\lambda_Z$ remains massless. In the closed tachyon condensation interpretation we fall into a critical 10d diagonal heterotic theory with 10d coordinates $X^i$, $Z$ (and $X$, $Y$ as extra supercritical multiplets, gapped with $\lambda_X$, $\lambda_Y$) and 32 left-moving fermions given by the 31 $\lambda_a$ and $\lambda_Z$. Amusingly, dimensions which are part of the physical 10d spacetime in one branch correspond to gapped supercritical dimensions in the others; similarly, left-moving fermions which are part of the 32 current algebra fermions of the 10d theory in one branch are massed up in the others.

Upon tachyon condensation the configuration results in a 6-branch junction of six 10d heterotic diagonal theories in 10d spacetimes bounded by a 9d junction, at which they join. Actually, the inclusion of the quantum corrections modifies this picture in exactly as in the type 0 junction in section \ref{sec:quantum} (because the relevant fields and interaction terms are exactly identical in both cases). Integrating out the massive fields in each of the branches turns the naive junction into a lightlike strongly coupled core, whose resolution is beyond the reach of worldsheet techniques. In what follows, we remain agnostic regarding the UV resolution of the strong coupling singularity, and simply explore the behavior of different theories and objects upon crossing them mostly from the perspective of the supercritical theory. Interestingly, our examples display junctions which are ready to admit a transmissible UV resolution, in the sense explained in the introduction.

\subsubsection{Other 6-branch junctions of 10d heterotic theories}
\label{sec:6branch-heterotic}

It is easy to introduce further $\IZ_2$ gaugings in the above 12d supercritical heterotic string configuration, compatible with the tachyon condensation, in order to produce junctions of other 10d heterotic theories. We now study the simplest possibility of $\IZ_2$ gaugings acting only on the fermions, leaving the bosons untouched (the introduction of $\IZ_2$ orbifolds acting on the bosons is postponed to sections \ref{sec:4-branch-junction-heterotic} and \ref{sec:3-branch-junction-heterotic}). A comprehensive discussion of such gaugings for supercritical heterotic theories can be found in Appendix \ref{sec:supercritical-heterotic-orbifold-no}, and the related 10d heterotic strings are discussed in Appendices \ref{sec:10dheterotic-diagonal}, \ref{sec:10d-nonsusy-heterotics}.

Let us consider the possible constructions which can be obtained for purely fermionic $\IZ_2$'s. First, we note that such $\IZ_2$ symmetries in supercritical heterotics  do not act on the right-moving fermions (see appendix \ref{sec:supercritical-heterotic-orbifold-no}). This means that, in order to be compatible with tachyon condensation, i.e. with the couplings (\ref{xyz-couplings-heterotic}), the $\IZ_2$ symmetries must leave $\lambda_X$, $\lambda_Y$ and $\lambda_Z$ invariant as well. This prevents quotienting the junction configuration by the $\IZ_2$ symmetries leading to the supercritical extension of the 10d supersymmetric theories or the 10d $SO(16)\times SO(16)$ theory, because these involve $\IZ_2$'s acting on 32 fermions (whereas here one can act only  on the 31 fermions $\lambda_a$). Hence it is not possible to use the above construction to provide junctions of those heterotic theories.

On the other hand, it is perfectly possible to perform the $\IZ_2$ gaugings leading to the supercritical versions of the $E_8\times SO(16)$ or the binary code heterotics. We leave their detailed discussion as an exercise for the interested reader, and rather turn to consider including $\IZ_2$ symmetries acting non-trivially on the bosons. As anticipated at the end of section \ref{sec:type0-junctions}, these have the advantage of reducing the number of branches in the junction.

\subsection{Adding a $\IZ_2$ orbifold to get 4-branch junctions}
\label{sec:4-branch-junction-heterotic}

In this section we start implementing extra $\IZ_2$ gaugings which flip some of the 2d bosons in the supercritical string, so they define orbifolds in spacetime, of the kind described in appendix \ref{sec:supercritical-heterotic-orbifold-yes}. In this section we consider including one such $\IZ_2$ action, which leads to 4-branch junctions. We will consider including further $\IZ_2$ orbifolds in section \ref{sec:3-branch-junction-heterotic}, leading to 3-branch junctions.

We thus focus on only one $\IZ_2$ gauging acting non-trivially on bosons, with possibly other $\IZ_2$ gaugings acting only on fermions. We are interested in actions compatible with the tachyon condensation process to the junction configuration, so we must ensure that the $\IZ_2$ actions are compatible with the interaction terms (\ref{xyz-couplings-heterotic}). We discuss the $\IZ_2$ acting on bosons first, and introduce possible fermionic $\IZ_2$ later on. Since our starting point is the 12d supercritical heterotic, the $\IZ_2$ must act on exactly 2 bosons, which without loss of generality we can take to be $X$, $Y$ (while $Z$ is invariant). This implies that it must also flip the right-moving fermions ${\tilde \psi}_X$, ${\tilde \psi}_Y$, and ${\tilde \psi}_Z$ is invariant (in other words, it acts as $(X,Y,Z)\to (-X,-Y,Z)$ on the full (0,1) multiplets). Then, enforcing the invariance of the fermionic terms in \eqref{xyz-couplings-heterotic}, we find that it must flip also the left-moving fermions $\lambda_X,\lambda_Y$, leaving $\lambda_Z$ invariant (so the $\IZ_2$ acts in non-chiral way in the (accidentally $(1,1)$) $X,Y,Z$ sector). In addition, we have to specify the action of the $\IZ_2$ on the remaining 31 left-moving fermions $\lambda_a$, which must correspond to flipping a multiple of 16 to achieve modular invariance. There are hence two possibilities, either flipping 16 out of the 31 fermions, or leaving all the 31 fermions invariant (corresponding to the $SO(16)\times SO(16+n)$  or $SO(32)\times SO(n)$ orbifolds in appendix \ref{sec:heterotic-orbifold-variants}, respectively). To recap, splitting the 31 fermions into 16 $\lambda_A$ and 15 $\lambda'_{A'}$, the two possible actions are
\beqa
&& g_{16}: (X,Y,Z) \to (-X,-Y,Z) \;\; , \;\; (\lambda_A,\lambda'_{A'},\lambda_X,\lambda_Y,\lambda_Z) \to (-\lambda_A,\lambda'_{A'}, -\lambda_X,-\lambda_Y,\lambda_Z)\nonumber \\
&& g_{0}: (X,Y,Z) \to (-X,-Y,Z) \;\; , \;\; (\lambda_A,\lambda'_{A'},\lambda_X,\lambda_Y,\lambda_Z) \to (\lambda_A,\lambda'_{A'}, -\lambda_X,-\lambda_Y,\lambda_Z)\, .\nonumber\\
\label{g16-g0}
\eeqa
The interpretation in spacetime as follows. The $Z$-branch is precisely at the fixed locus of the $\IZ_2$ orbifold action, so the resulting 10d theory after tachyon condensation is that arising from the 12d on after imposing the orbifold projection. In other words, in the $Z$-brane we get the 10d  $E_8\times SO(16)$ heterotic theory if the orbifold action is $g_{16}$, or the 10d non-supersymmetric diagonal heterotic theory if the orbifold action is $g_0$. In the $X$-branch, the $\IZ_2$ orbifold acts by mapping the $X<0$ region to the $X>0$, so we simply keep on of them in the quotient and do not impose any projection. The result is that, upon tachyon condensation, the $X>0$ branch supports a 10d non-supersymmetric diagonal heterotic theory. Similarly for the $Y$-branch, so upon tachyon condensation the $Y>0$ branch also supports a 10d non-supersymmetric diagonal theory as well.

Overall, the construction leads to a 4-branch junction of either four 10d diagonal non-supersymmetric heterotic strings in 10d spacetimes with a 9d boundary at which they are all joined, or a 4-branch junction of two 10d diagonal and two 10d $E_8\times SO(16)$ heterotic string theories. The fact that we have obtained precisely two 10d $E_8\times SO(16)$ heterotic theories is natural from the perspective of chirality. These theories contain a set of chiral fermions, so it is not easy to define boundary conditions for them. In the above configuration this is solved, as the chiral fermions of the theory at the $Z<0$ branch can simply cross the 9d junction and continue as chiral fermions of the theory at the $Z>0$ branch. This is related to the fact that one can add a superpotential invariant under the orbifold action, deforming the junction into a recombination of the $X$- and $Y$-branches, with the $Z$-branch ending up a a disconnected $E_8\times SO(16)$ heterotic theory in full (rather than half) 10d spacetime.

Let us now consider the possibility of adding extra $\IZ_2$ gaugings acting only on the fermions. For concreteness, we consider starting from the junction obtained from $g_{16}$. In order to be compatible with the tachyon condensation, namely the interactions (\ref{xyz-couplings-heterotic}), they must be of the kind considered in section \ref{sec:6branch-heterotic}. In particular it is not possible to introduce the gaugings reproducing the 10d supersymmetric heterotics, as they require gaugings acting on 32 fermions, which are not compatible with (\ref{xyz-couplings-heterotic}). On the other hand, we may introduce gaugings flipping 16 left-moving fermions. These need not be the 16 $\lambda_A$ flipped by $g_{16}$, although one must ensure that their products (and the products with $g_{16}$) still act by flipping multiples of 16 fermions out of the 32 fermions in the 10d theory. This precisely leads to the classification of $\IZ_2$ actions via the binary code labeling of current algebra fermions, explained in section \ref{sec:10d-nonsusy-others}. 

For illustration, let us spell it out for the case of just one extra $\IZ_2$ besides $g_{16}$. Consider splitting the 31 fermions $\lambda_a$ into 8+8+8+7 denoted $\lambda_{A_1}$, $\lambda_{A_2}$, $\lambda'_{A'_1}$, $\lambda'_{A'_2}$. Then we take $g_{16}$ to act as in (\ref{g16-g0}), by flipping $\lambda_{A_1}$, $\lambda_{A_2}$ (besides  $\lambda_X$, $\lambda_Y$ and the supermultiplets $X$, $Y$). We take the generator $\theta_{16}$ of the further $\IZ_2$ to simply flip the 16 fermions $\lambda_{A_1}$, $\lambda'_{A'_1}$, leaving all other fields invariant. Their product $g_{16}\theta_{16}$ hence flips the 32 fermions $\lambda_{A_2}$, $\lambda'_{A'_1}$ (together with $\lambda_X$, $\lambda_Y$ and the supermultiplets $X$, $Y$). After the tachyon condensation, we obtain the following pattern. The $Z$-branch is at the orbifold fixed locus of $g_{16}$ and also experiences the projection by $\theta_{16}$, hence it becomes the 10d binary code heterotic theory with $\IZ_2\times \IZ_2$ action; this has gauge group $[E_7\times SU(2)]^2$ and has a chiral spectrum whose discussion we skip. On the other hand, the action of $g_{16}$ exchanges the $X<0$ and $X>0$ regions of the $X$-branch, so we keep one copy and do not impose this projection, but we must still impose the $\theta_{16}$ projection. This gives a 10d $E_8\times SO(16)$ heterotic theory in the $X>0$ branch. Finally, there is similarly a 10d $E_8\times SO(16)$ heterotic theory in the $Y>0$ branch. 

The final result is a 4-branch junction of two 10d $E_8\times SO(16)$ heterotic theories and two 10d $[E_7\times SU(2)]^2$ heterotic theories. Again, even though each branch supports a chiral 10d theory, they are paired up so that chiral fields can cross from one theory to the corresponding partner theory across the junction. Clearly a similar pattern holds upon introducing additional binary code $\IZ_2$ gaugings, which we refrain from discussing further. We instead turn to a more interesting configuration leading to 3-branch junctions, which display a bouquet with non-trivial chiral flow across the junction.

\subsection{A heterotic chiral bouquet using $\IZ_2\times\IZ_2$ orbifolds}
\label{sec:3-branch-junction-heterotic}

In this section we consider including additional $\IZ_2$ orbifolds acting on the bosons, and use them to build 3-branch junctions. We will interestingly find 3-branch junctions of 10d chiral heterotic theories, for instance of three $E_8\times SO(16)$ heterotics. This is a bouquet configuration with non-trivial chiral flow across the junctions.

\subsubsection{3-branch junction of 3 $E_8\times SO(16)$ heterotic theories}
\label{sec:3-branch-e8-so16}

We focus on introducing a $\IZ_2\times\IZ_2$ orbifold acting non-trivially on the bosons (there may be additional $\IZ_2$ gaugings acting only on the fermions, which will be discussed later on). In order for the orbifold to be compatible with the tachyon condensation, its actions must preserve the interactions (\ref{xyz-couplings-heterotic}). We can then consider the elements of the orbifold group to be actions of the kind considered in section \ref{sec:4-branch-junction-heterotic}. For concreteness we will focus on actions of the kind $g_{16}$, which lead to chiral 10d theories more interesting than the 10d diagonal heterotics arising for $g_0$. In order to ensure that the actions on the left-moving fermions are compatible, we choose a (binary code motivated) splitting of the 31 fermions $\lambda_a$ into 8 $\lambda_{A_1}$, 8 $\lambda_{A_2}$, 8 $\lambda'_{A'_1}$ and 7 $\lambda'_{A'_2}$.
A simple choice satisfying the compatibility criteria is
\beqa
& g_{16,Z}: &(X,Y,Z) \to (-X,-Y,Z) \;\; , \label{z2z2-heterotic-1} \\
&& (\lambda_{A_1},\lambda_{A_2},\lambda'_{A'_1}, \lambda'_{A'_2},\lambda_X,\lambda_Y,\lambda_Z) \to (-\lambda_{A_1},-\lambda_{A_2},\lambda'_{A'_1}, \lambda'_{A'_2}, -\lambda_X,-\lambda_Y,\lambda_Z)\nonumber \\
& g_{16,X}: &(X,Y,Z) \to (X,-Y,-Z) \;\; ,  \label{z2z2-heterotic-2}\\
&& (\lambda_{A_1},\lambda_{A_2},\lambda'_{A'_1}, \lambda'_{A'_2},\lambda_X,\lambda_Y,\lambda_Z) \to (\lambda_{A_1},-\lambda_{A_2},-\lambda'_{A'_1}, \lambda'_{A'_2}, \lambda_X,-\lambda_Y,-\lambda_Z)\nonumber \\
& g_{16,Y}: &(X,Y,Z) \to (-X,Y,-Z) \;\; ,  \label{z2z2-heterotic-3}\\
&& (\lambda_{A_1},\lambda_{A_2},\lambda'_{A'_1}, \lambda'_{A'_2},\lambda_X,\lambda_Y,\lambda_Z) \to (-\lambda_{A_1},\lambda_{A_2},-\lambda'_{A'_1}, \lambda'_{A'_2}, -\lambda_X,\lambda_Y,-\lambda_Z)\, ,\nonumber 
\eeqa
where $X,Y,Z$ denote the full $(0,1)$ multiplets including the bosons and the right-moving fermions. We note that each action in the $\IZ_2\times \IZ_2$ orbifold is of the kind $g_{16}$ in (\ref{g16-g0}), acting on different sets of 2d fields (with the invariant one indicated as subindex). It is easy to check that the above action, when restricted to the (accidentally $(1,1)$) $X,Y,Z$ sector, is the only possible $\IZ_2\times\IZ_2$ symmetry preserving (\ref{xyz-couplings-heterotic}) and acting non-trivially on the bosons. Clearly, there are other possible choices of actions on the remaining 31 left-moving fermions, as we discuss later on. For the moment we proceed with the above one, which leads to the most interesting configuration.

The geometric action on the bosons implies that the orbifold fixed point of the action $g_{16,Z}$, given by $X=Y=0$ (i.e. the $Z$-branch after tachyon condensation), has its $Z>0$ and $Z<0$ parts mapped to each other by both $g_{16,X}$ and $g_{16,Y}$, so we keep one of the copies (e.g. $Z>0$) and do not impose those projections. Hence, the resulting twisted sector at this orbifold fixed point is (before tachyon condensation) that of the $SO(16)\times SO(16+n)$ theory studied in section \ref{sec:heterotic-orbifold-variants}. This is just like in section \ref{sec:4-branch-junction-heterotic}, but on the $Z>0$ branch alone, due to the extra orbifold actions. Upon tachyon condensation the junction contains a $Z>0$ brane supporting a 10d $E_8\times SO(16)$ heterotic theory, on 10d spacetime bounded by the 9d junction at $Z=0$. 

\begin{figure}[htb]
\begin{center}
\includegraphics[scale=.3]{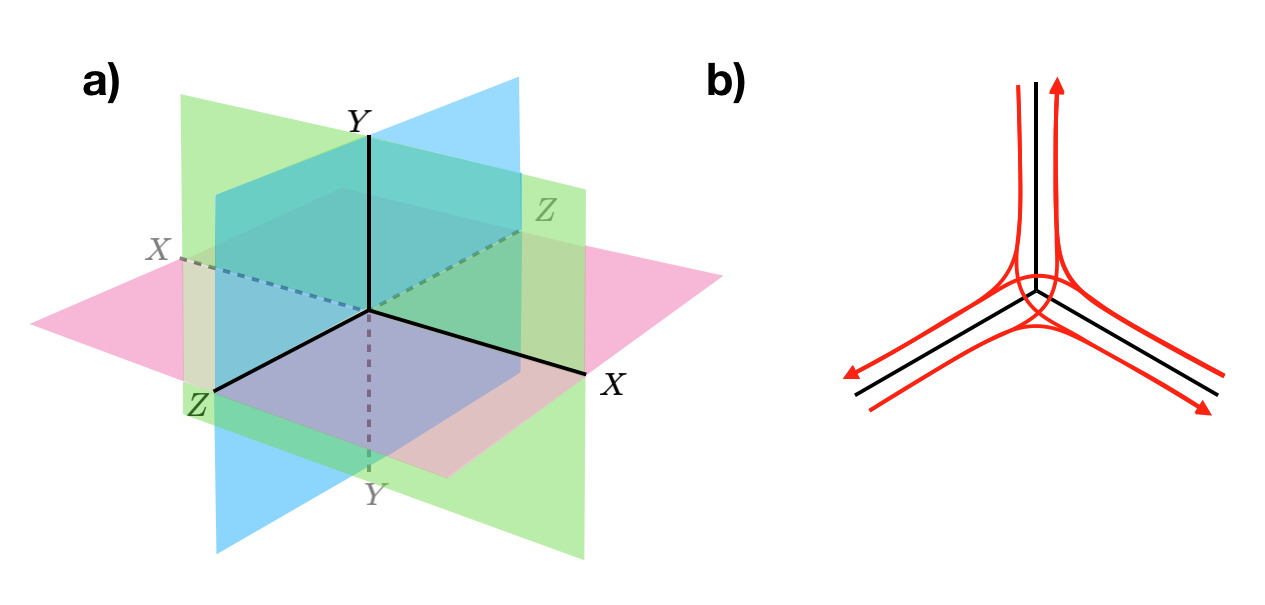}
\caption{\small a) The 3-branch junction of three 10d $E_8\times SO(16)$ heterotic theories. The colored planes are fixed under the orbifold actions, and the dashed lines correspond to the images of some of the branches under them. b) The no-trivial flow of chiral fields across the junction, resulting in a bouquet configuration.}
\label{fig:flow-heterotic}
\end{center}
\end{figure}

Clearly, we have a similar pattern for the $X$- and $Y$-branches. Before tachyon condensation we have three orbifold fixed loci supporting twisted sectors of the $SO(16)\times SO(16+n)$ theory, albeit with the gauge factors on each branch arising from different subsets of the left-moving fermions. Upon tachyon condensation, the final configuration is a 3-branch junction of three $E_8\times SO(16)$ heterotic theories in 10d spacetimes with a 9d boundary, at which they join, see Figure \ref{fig:flow-heterotic}a.

%%%%%%%%%%%%
%%%%%%%%%%%%

This is a remarkable configuration, because the chiral fermions of the different branches must have non-trivial boundary conditions at the junction, determining whether they are gapped at the junction, or whether (and how) they propagate from one branch to another (or to several of them). We will discuss this in the next section. As a spoiler, we anticipate that the junction actually supports a non-trivial chiral flow, so it describes a bouquet. This is correlated with the fact that the orbifold projections remove the possibility of adding the deformations to the superpotential which would account for the recombination of the branches. This implies that the bouquet configuration is particularly robust due to the chiral flow it supports.

\medskip

Let us finally briefly mention that it is possible to consider variants of the above construction, for instance by replacing the actions of type $g_{16}$ by actions of type $g_0$ in (\ref{z2z2-heterotic-1})-(\ref{z2z2-heterotic-3}), while maintaining their action on $\lambda_X$, $\lambda_Y$, $\lambda_Z$ and the $(0,1)$ multiplets $X,Y,Z$. For instance we may replace the three elements of $\IZ_2\times \IZ_2$ by similar elements $g_{0,Z}$, $g_{0,X}$, $g_{0,Y}$ leaving the 31 left-moving fermions invariant. This results in a 3-branch junction of three 10d diagonal heterotic theories. We can also maintain $g_{16,Z}$, and replace $g_{16,X}\to g_{0,X}$, and consequently modify $g_{16,Y}$ by having it act on the 31 left-moving fermions as $g_{16,Z}$. This results in a 3-branch juntion of two 10d $E_8\times SO(16)$ theories and one 10d diagonal heterotic theory. In these other configurations, chirality is either absent, or its flow is a simple propagation between two identical theories across the junction. We refrain from considering further these other less interesting constructions.

\subsubsection{Junction conditions for the 3-brach bouquet of $E_8\times SO(16)$ theories}
\label{sec:junction-conditions-heterotic}

We now discuss the fate of the 10d fields of the $E_8\times SO(16)$ theories on the different branches as they reach the junction. This can be thought of as describing the junction conditions for the different fields. This question is particularly interesting for the 10d chiral fermions in the $E_8\times SO(16)$ theories on the three branches. One possibility is that these fields are non-trivially gapped at the junction (which operates as a boundary for each branch), hence requiring a non-trivial strong coupling symmetric mass generation mechanism to gap the non-anomalous set of chiral  fermions; such mechanisms are however not known in 10d theories (see \cite{Angius:2024pqk} for the construction of boundary configurations for 6d and 4d chiral gravitational theories, and e.g. \cite{Razamat:2020kyf,Tong:2021phe,Wang:2022ucy} for 2d and 4d QFT realizations of symmetric mass generation). A simpler possibility is that the chiral fermions of one branch move on to another branch, or several of them, hence defining junction conditions for the different fields. 

In other words, the junction must support either a strong coupling symmetric mass generation mechanism (of still unknown nature), or a consistent transmission of the chiral fields. Our junctions do have a strongly coupled core, as already explained, so we cannot exclude the presence of symmetric mass generation, but its nature would be beyond the reach of the worldsheet techniques we are using. On the other hand, we will now show that there are easily derived junction conditions, which render the theory ready to admit a transmissible UV completion of the junction.

The junctions conditions for chiral fields are largely constrained by the topological protection these fields enjoy. However, there are some interesting subtleties arising from the detailed worldsheet structure in our construction. For instance, recall from section \ref{sec:10d-nonsusy-e8so(16)} that the 10d chiral fields in each $E_8\times SO(16)$ theory are fermions transforming as $({\bf 8}_s, {\bf 128})+({\bf 8_c},{\bf 128'})$ of the $SO(8)\times SO(16)$ Lorentz and gauge groups. Hence, one may be tempted to claim that, as a chiral fermion in the $({\bf 8}_s, {\bf 128})$ of e.g. the $X$-branch crosses the junction, it turns into a chiral fermion in the $({\bf 8}_s, {\bf 128})$ of e.g. the $Y$-branch. However, it is clear from our construction that this is not possible. Indeed, in the supercritical realization of our junction, the $SO(16)$ in the two different branches arises from different subsets of left-moving fermions, as well as the Lorentz $SO(8)$ of the two different branches arises from different subsets of right-moving fermions (corresponding to the fact that the 10d spacetime dimensions in the two branches arise from a common 9d spacetime plus the directions $X$ or $Y$, for one or the other branch). 

A systematic way to derive the junction conditions is to consider the possible non-zero correlators of vertex operators of fields in the different branches, which can be computed in the underlying supercritical theory, which is a common arena for all the branches. Upon tachyon condensation, the information related to the (branch-dependent) supercritical directions is erased by the possible insertion of tachyon background vertex operators in the correlator, resulting in a set of allowed correlators for 10d fields in different branches. In practice, this amounts to checking selection rules among 10d fields in different branches, including the information of the embedding of their quantum numbers in the supercritical description. In what follows, we provide a heuristic derivation of the result of this exercise, to determine the junction conditions for the chiral fermions (the chiral flow of the bouquet), leaving a more detailed description for Appendix \ref{app:correlators}. The punchline is that a chiral fermion in a given branch can propagate as a chiral fermion in a second branch if it simultaneously emits an $E_8$ gauge boson in the third branch. 

The key idea is that, in order to compare the symmetries of the different branches, we must resort to the common symmetry preserved at the junction. Considering the splitting of the 34 left-moving fermions into 8+8+8+7+1+1+1 (namely the 8 $\lambda_{A_1}$, 8 $\lambda_{A_2}$, 8 $\lambda'_{A'_1}$, 7 $\lambda'_{A'_2}$, and $\lambda_X$, $\lambda_Y$, $\lambda_Z$), we have a preserved common $SO(8)_X\times SO(8)_Y\times SO(8)_Z\times SO(7)$ symmetry at the junction. The subindices are convenient because e.g. on the $X$-branch, the 10d $SO(16)$ gauge group (dubbed $SO(16)_X$ in the following) arises from the enhancement of $SO(8)_X\times SO(7)$ together with $\lambda_X$, and similarly for the other branches. Note also that the 10d $E_8$ on the $X$-branch (which we denote by $(E_8)_X$) arises from the $SO(8)_Y\times SO(8)_Z$ (together with $\lambda_Y$, $\lambda_Z$, which disappear from the $X$-branch upon tachyon condensation), and similarly for the other branches. There is a similar split of the Lorentz $SO(8)$ groups in each branch (dubbed $SO(8)'_X$, etc), which leads to a common $SO(7)$, see Appendix \ref{app:correlators} for details.

Consider now a chiral spinor $({\bf 128},{\bf 8}_s)_X$ of $SO(16)_X\times SO(8)'_X$ of the gauge and Lorentz groups in the $X$-branch, which transforms as a bi-spinor $({\bf 8}_{s,X},{\bf 8})+({\bf 8}_{c,X},{\bf 8})$ under the gauge symmetry $SO(8)_X\times SO(7)$, and a spinor ${\bf 8}$ of the Lorentz $SO(7)$. We want to check if this can propagate into the $Y$-branch as a spinor $({\bf 128}_Y,{\bf 8}_s)_Y$. This transforms as a bi-spinor $({\bf 8}_{s,Y},{\bf 8})+({\bf 8}_{c,Y},{\bf 8})$ under the gauge symmetry $SO(8)_Y\times SO(7)$, and a spinor ${\bf 8}$ of the Lorentz $SO(7)$. This is possible if it simultaneously emits a state in the $Z$-branch, transforming as a bi-spinor $({\bf 8}_{s,X},{\bf 8}_{s,Y})+({\bf 8}_{c,X},{\bf 8}_{c,Y})$ under the gauge symmetry $SO(8)_X\times SO(8)_Y$, and as a product of two $SO(7)$ Lorentz spinors. This are precisely the quantum numbers of an $(E_8)_Z$ gauge boson (in the ${\bf 128}$ spinor representation of the underlying $SO(16)\supset SO(8)_X\times SO(8)_Y$) on the $Z$-branch.

Clearly, a similar analysis can be carried out for the fermions in the $({\bf 128'},{\bf 8}_c)$ of the $X$-branch, turning into the $({\bf 128'},{\bf 8}_c)$ in the $Y$-branch, by emitting an $E_8$ gauge boson in the $Z$ branch. Note that, by the permutation symmetry of the branches, we have a similar pattern for other combinations of spinors in any two other branches and a gauge boson in the third. We also note that it is not possible for a $({\bf 128},{\bf 8}_s)$ in one branch to turn into a $({\bf 128'},{\bf 8}_c)$ in another, since there is no 
$E_8$ gauge boson with the correct quantum numbers in the third. The non-trivial chiral flow for this bouquet is shown in Figure \ref{fig:flow-heterotic}b. We again refer to Appendix \ref{app:correlators} for additional details, and leave the application of the ideas therein to junctions conditions for the non-chiral fields as an exercise for the interested reader.

\section{Grand finale: the IIB-$SO(32)$-$USp(32)$-$U(32)$ bouquet}
\label{sec:bouquet}

The configurations in the previous sections provide an interesting class of multi-branched junctions of several 10d string theories, including bouquets with non-trivial flows of chiral fields among the different branches. However, we have not encountered yet any bouquet configuration involving any of the 10d supersymmetric string theories. In this section we fill this gap and provide the construction of a bouquet including  10d supersymmetric strings in some of their branches, with a non-trivial chiral flow among multiple branches. This will arise from a construction involving unoriented strings.

Hence, in this section we explore the realization of junctions involving unoriented string theories, such as type I. This can be achieved by implementing an orientifold projection on the type 0 and type II junctions considered in section \ref{sec:type0-typeii}. We provide an explicit example of a remarkable 4-branch bouquet involving the four non-tachyonic 10d descendants of type 0B theory: type IIB, type I, the 10d non-supersymmetric $USp(32)$ theory \cite{Sugimoto:1999tx}, and the 10d non-supersymmetric $U(32)$ 0B orientifold (sometimes called 0'B theory) \cite{Sagnotti:1995ga,Sagnotti:1996qj}. We refer to this beautiful configuration as the IIB-$SO(32)$-$USp(32)$-$U(32)$ bouquet, or IIB bouquet, for short, with a 10d type IIB stem base at whose 9d boundary the three individual type I, $USp(32)$ and 0'B stems join.

\subsection{The orientifold group}
\label{sec:orientifold-group}

The systematic construction of orientifolds of supercritical string configurations has not been carried out in the literature, and is beyond the scope of thiw work. Hence we will simply provide the necessary ingredients for the construction of the configuration.

The strategy is to start with the 6-branch junction of type 0B theories in section \ref{sec:type0-junctions} and to impose an orbifold projection leading to type IIB branches, and an extra orientifold projection. We thus take the 12d type 0B theory, with 7 (1,1) multiplets with bosons $X^i$ and fermions $\psi^i, {\tilde\psi}^i$ (for left- and right-movers, respectively) and 3 extra multiplets with bosons $X,Y,Z$, and fermions $\psi_X,\psi_Y,\psi_Z$ (and their tilded versions), and superpotential (\ref{supo-xyz}), leading to the interactions (\ref{couplings-xyz-type0}) in components. One may also try to consider the $Y^3$ deformation (\ref{y-cube-deformation}) to reduce the number of branches as in section \ref{sec:ii-ii}, but we will show this is not compatible with our orientifold projection.

The discussion of the orbifold projection was discussed in section \ref{sec:ii-ii}. We consider gauging by the $\IZ_2$ generated by 
\beqa
g_{II}=R_Y(-1)^{{\tilde F}_Y}(-1)^{F_X+F_Z+F_i}\,.
\label{g-II}
\eeqa
The explicit action is given in (\ref{orbifold-typeii}). Let us recall that, in the quotient, the $X$- and $Z$-branches are mapped to themselves, so they experience the orbifold projection and become 10d type IIB theories; on the other hand, the $Y<0$ branch is exchanged with the $Y>0$ branch, so we keep one of them with no projection. We end up with a 5-branch junction, with four 10d type IIB theories and one 10d 0B theory.

We now wish to introduce an orientifold action, to projects some of the 10d theories to unoriented descendants thereof. We want to focus on orientifolds by $g_{\Omega}=\Omega h$, where $\Omega$ is worldsheet parity and $h$ includes a geometric action (i.e. on the bosons) and an action on the fermions. For the geometric action, we focus on $R_{X}R_Y:(X,Y,Z)\to (-X,-Y,Z)$, as it leads to an interesting structure: It maps the $X<0$ branch to the $X>0$ branch, so that we keep only one copy (e.g. the $X>0$ branch) and impose no projection (hence, it still supports a 10d type IIB theory); on the other hand, the coordinate $Z$ is invariant, so that the two type IIB branches $Z<0$ and $Z>0$ suffer the projection and become unoriented IIB theories (namely, type I or $USp(32)$ theories, as we specify later on).

We now must fix the action on the fermions. In order to recover 10d Poincar\'e invariant theories on the $Z$ branch, the quotient should be just by $\Omega$ times operations that act only on the supercritical coordinates from the perspective of the $Z$ branch, namely bosons and fermions associated to the $X$  and $Y$ direction. We in fact had already fixed this for the bosons, so we now demand it for the fermions. A natural guess, suggested by $(1,1)$ supersymmetry, is the fermionic action is $(-1)^{F_X+F_Y}(-1)^{{\tilde F}_X+{\tilde F}_Y}$. The total action is therefore
\beqa
g_{\Omega}=\Omega R_XR_Y (-1)^{{\tilde F}_X+{\tilde F}_Y}(-1)^{F_X+F_Y}\, .
\label{g-omega}
\eeqa
This action is compatible with the interactions (\ref{couplings-xyz-type0}), namely the interaction terms pick up an overall sign, which cancels against the sign flip (due to $\Omega$) of the integration measure $d\sigma dt$ in the 2d action. On the other hand, the interactions arising from the cubic deformation (\ref{y-cube-deformation}) --or in fact, any monomial deformation in $Y$-- are not invariant under the above orientifold action, so the $Y$-branch cannot be lifted. This will receive a simple spacetime interpretation in section \ref{sec:bouquet-flow}.

Since we are quotienting by $g_{II}$ and $g_\Omega$, the orientifold group also includes the element $g_{\Omega}g_{II}$, given by
\beqa
&g_{\Omega}g_{II}&=\Omega R_XR_Y (-1)^{{\tilde F}_X+{\tilde F}_Y}(-1)^{F_X+F_Y}R_Y(-1)^{{\tilde F}_Y}(-1)^{F_X+F_Z+F_i}=\nonumber\\
&&= \Omega R_X(-1)^{{\tilde F}_X}(-1)^{F_Y+F_Z+F_i}\, .
\label{combo}
\eeqa
Invariance of the interactions (\ref{couplings-xyz-type0}) follows automatically. The action flips the $X$ coordinate, so it exchanges the $X<0$ and the $X>0$ branch, hence there is no orientifold projection on it, and still supports a 10d type IIB theory. Each of the $Z<0$ and $Z>0$ branches are mapped to themselves (as expected, because they were mapped to themselves by $g_{II}$ and $g_\Omega$), so they still support 10d orientifolds of 10d type IIB theory. Finally, each of the $Y<0$ and $Y>0$ are mapped to themselves so each supports a 10d orientifold of the type 0B theory (actually, since both branches are related by $g_{II}$, we simply get one copy, say $Y>0$), to be specified next. 

\subsection{The 10d orientifold theories in the IIB bouquet}
\label{sec:orientifold-theories}

We now turn to determine the specific orientifold theories we get on the $Y>0$ branch and the $Z<0$ and $Z>0$ branches. The 10d orientifold action on the $Y$ branch can be read from $g_\Omega g_{II}$ in (\ref{combo}) by `forgetting' about the action on fields associated to the $X$ and $Z$ directions. We explicitly have
\beqa
g_{\Omega}g_{II}\; \rightarrow \;\Omega (-1)^{F_Y+F_i}= \Omega (-1)^{F_{10d,Y}}\, ,
\eeqa
where $F_{10d,Y}=F_i+F_Y$ is left-moving worldsheet fermion number for the 10d critical theory obtained on the $Y$ branch after tachyon condensation. In more standard 10d orientifold notation, the orientifold action is $\Omega (-1)^{F_L}$, where $F_L$ is the left-moving worldsheet fermion number of the 10d theory. This corresponds to the (closed string sector of the) 0'B theory (sometimes also called $U(32)$ 0B orientifold or $\Omega (-1)^{F_L}$ 0B orientifold) \cite{Sagnotti:1995ga,Sagnotti:1996qj}. We recall that this orientifold acts on the 0B theory by removing the closed tachyon, and exchanging the two sets of RR fields, leading to a RR sector with one 0-form, one 2-form and one self-dual 4-form.

The RR tadpole cancellation in this 10d theory requires the introduction of D9-branes, as follows (see \cite{Sagnotti:1995ga,Sagnotti:1996qj}, and e.g. \cite{Dudas:2000sn,Angelantonj:2002ct,Raucci:2024fnp} for more details). There are two different kinds of D9-branes (denoted D9$_1$ and D9$_2$ in what follows), charged under either of the two RR 10-forms. Hence, they have tensions and charges $(T;Q_1,Q_2)_{{\rm D9}_1}=(+1;+1,0)$ and $(T;Q_1,Q_2)_{{\rm D9}_2}=(+1;0,+1)$, in suitable units. The orientifold introduced by $\Omega (-1)^{F_L}$ carries no tension (i.e. the closed channel Klein bottle includes no exchange of NSNS fields), but has charge $-32$ under both 10-forms, namely $(T;Q_1,Q_2)_{\rm O9}=(0;-32,-32)$. Cancellation of RR tadpoles requires the introduction of 32 D9$_1$- and 32 D9$_2$-branes. The orientifold acts by exchanging them, so the gauge group is $U(32)$\footnote{This is ultimately broken to $SU(32)$ by a St\"uckelberg coupling with the 8-form dual of the RR 0-form, necessary for the Greenn-Schwarz anomaly cancellation mechanism.}, and the $9_19_2$ and $9_29_1$ sectors lead to equal chirality fermions in the $\asymm+\antiasymm$ of $U(32)$. 

In our 12d supercritical setup we have an identical situation, with the proviso that the RR 10-form tadpoles arise in crosscaps of $g_{\Omega}g_{II}$ in (\ref{combo}). These RR tadpoles are again cancelled by a set of D9-branes (which are not spacetime filling in 12d, but are so in the fixed 10d $Y$-branch). Hence, at $Y>0$ we have a 0'B theory, with the above described spectrum. The discussion of  how it behaves as we move across the junction and into the other branches will be further discussed in section \ref{sec:bouquet-flow}.

Let us now turn to the two theories on the $Z<0$ and $Z>0$ branches. As explained above, they correspond to 10d orientifolds of the underlying IIB theory (after the quotient by $g_{II}$). There are two such possible theories, given by type I and the non-supersymmetric $USp(32)$ theories. The former can be described as the introduction of an O9$^-$-plane in the underlying IIB theory plus 32 D9-branes to cancel the RR 10-form tadpole. The latter can be described as the introduction of an anti-O9$^+$-plane\footnote{The $USp(32)$ theory is often described using an O9$^+$-plane and 32 anti-D9-branes. We use an equivalent description more convenient for our purposes.} (denoted by ${\ov{{\rm O9}^+}}$) in the underlying IIB theory plus 32 D9-branes to cancel the RR 10-form tadpole (in this case, both the ${\ov{{\rm O9}^+}}$-plane and the D9-branes carry positive tension, so the NSNS tadpole does not cancel, resulting in a dilaton tadpole).

In principle one can use the detailed consistency of the orientifold and orbifold actions to fix the discrete choices determining which combination of theories can be obtained in the $Z<0$ and $Z>0$ branches. The result can however be easily guessed as follows. The 0'B theory in the $Y>0$ branch has two sets of 32 D9-branes, so it is natural to guess that they become the 32 + 32 D9-branes in the $Z<0$ and $Z>0$ branches. Since in the latter the orientifold action maps each set of D9-branes to itself, they are related to combinations of 0'B D9-branes given by eigenstates of the 0'B orientifold action. The latter exchanges the D9$_1$- and D9$_2$-branes, so the orientifold acts with a $+1$ sign on 32 D9-brane combinations and $-1$ on the other 32 D9-branes. This implies that in the $Z$-branch, the Moebius strip amplitude in the $Z<0$ and $Z>0$ branches have opposite signs, implying that the corresponding D9-branes suffer opposite $SO/USp$ projections. Hence, we conclude that one of the branches (say the $Z<0$ branch) supports a type I theory, with gauge group $SO(32)$, while the other (the $Z>0$ branch) supports a 10d $USp(32)$ theory. 

One can reach the same conclusion by noting that the O9-plane of the 0'B theory has tension and charges $(T;Q_1,Q_2)_{\rm O9}=(0;-32,-32)$, so it can be formally regarded as a combination of an O9$^-$-plane (with tension $-32$ and charge $-32$) and an ${\ov{{\rm O9}^+}}$-plane (with tension $+32$ and charge $-32$). Hence the 0'B O9-plane in the $Y>0$ branch can split into a type I O9-plane and an $USp(32)$ ${\ov{{\rm O9}^+}}$-plane in the $Z<0$ and $Z>0$ branches, respectively. 

\medskip

\begin{figure}[htb]
\begin{center}
\includegraphics[scale=.3]{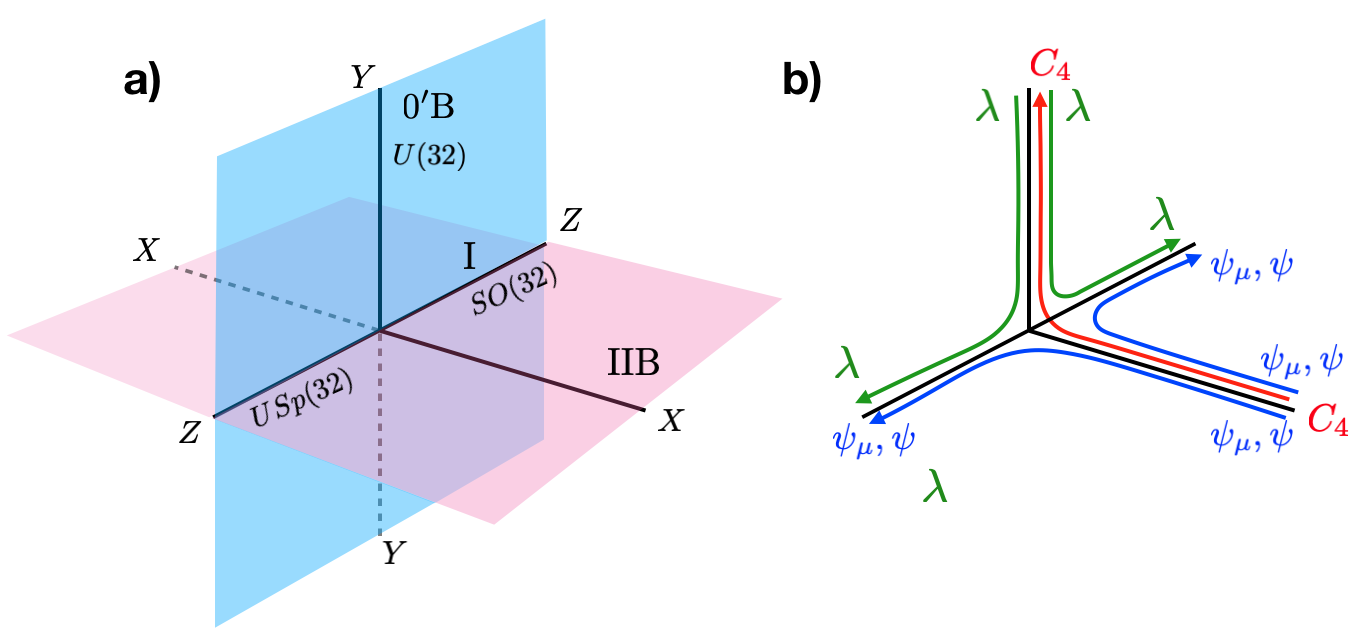}
\caption{\small a) The bouquet junction, with the type IIB, type I, $USp(32)$ and 0'B theory. The colored planes in pink and blue are fixed under the orbifold and orientifold actions, respectively, and the dashed lines correspond to the images of some of the branches under them.
b) The flow of chiral fields, including the gravitino/dilatinos (in blue), the RR 4-form (in red) and the fermions in two-index antisymmetric representation or its conjugate (in green).}
\label{fig:bouquet}
\end{center}
\end{figure}

This concludes the construction of the IIB bouquet configuration, which beautifully combines the four 10d non-tachyonic descendants of type 0B theory, or alternatively gives the branching of a 10d type IIB theory into its three 10d non-tachyonic unoriented cousins, see Figure \ref{fig:bouquet}a. In the following section we discuss the 10d spacetime spectrum on the various branches, focusing on the flow of 10d chiral fields among them.

%%%%%%%%%%%%
%%%%%%%%%%%%

\subsection{Junction conditions: chiral flow in the IIB bouquet}
\label{sec:bouquet-flow}

We now discuss the junction conditions among the theories in the different branches. In particular, we are interested in the 10d chiral spectrum of these theories, to check whether the non-anomalous combinations of chiral fields are gapped at the junction, or whether they propagate between different branches. The techniques to study this in detail are similar to those used in section \ref{sec:junction-conditions-heterotic}, but the discussion in this case is simpler and follows from anomaly matching across the junctions. We will find that all the chiral fields have a simple way to arrange into a consistent chiral flow in the bouquet. 

Let us begin by recalling the chiral spectrum of the 10d type IIB, type I, $USp(32)$ and 0'B theories: 

\begin{itemize}

\item The type IIB theory contains the two 10d $\NN=2$ gravitino/dilatinos and one self-dual 4-form. This field content is anomaly-free and no Green-Schwarz mechanism is required.

\item The type I theory contains the 10d $\NN=1$ gravitino/dilatino (of same chirality as the IIB parent theory) and the 10d $\NN=1$ gauginos in the 496-dimensional representation $\asymm$ of $SO(32)$. The anomaly polynomial factorizes as $I_{12}=X_4X_8$ and the anomaly is cancelled by a Green-Schwarz mechanism mediated by the RR 2-form.

\item The $USp(32)$ theory contains the 10d $\NN=1$ gravitino/dilatino (of same chirality as the IIB parent theory), and a (non-supersymmetric) set of 496 fermions in the $\asymm+1$ (reducible) representation of $USp(32)$. We loosely refer to the latter fermions as `gauginos' by analogy with the corresponding fields in type I. The anomaly polynomical is identical to that of type I theory, and so factorizes as $I_{12}=X_4X_8$ and the anomaly is cancelled by a Green-Schwarz mechanism mediated by the RR 2-form.

\item The 0'B theory contains the RR self-dual 4-form and two sets of same chirality fermions in the $2\times 496$-dimensional representation $\asymm+\antiasymm$ of $SU(32)$. The anomaly polynomial is a sum of factorized contributions $I=X_2X_10+X_4X_8+X_6X_6$, and is cancelled by Green-Schwarz mechanisms mediated by the RR 0-, 2- and 4-forms \cite{Sagnotti:1995ga,Sagnotti:1996qj}.

\end{itemize}

Given these ingredients, the structure of the resulting flow of chiral fields at the junction is clear (and certainly amusing), see Figure \ref{fig:bouquet}b. The two 10d gravitino/dilatinos of the IIB branch split into the 10d gravitino/dilatinos of the type I and $USp(32)$ branches. On the other hand, the RR 4-form in the IIB branch turns into the 4-form in the 0'B branch. Finally, the two fermions in the 0'B branch turn into the gauginos of the type I and $USp(32)$ branches. In this regard, we should note that both the $\asymm$ and the $\antiasymm$ of $SU(32)$ turn into the $\asymm$ of $SO(32)$ and the $\asymm+1$ of $USp(32)$, when these symmetries are regarded as subgroups of $SU(32)$, so there is no actual mismatch of gauge representations.

This non-trivial chiral flow explains why it was not possible to remove the $Y$-branch via a cubic deformation of the superpotential (which was forbidden by the orientifold projection). This would have removed the 0'B branch from the bouquet, leaving no possibility to achieve a consistent chiral flow. If is easy to check that any other deformation of the junction by extra superpotential terms is similarly forbidden by the symmetries we have gauged to achieve our construction. This is a nice illustration of the link between the robustness of the bouquet and the non-trivial chiral flow it supports.

A further question about the non-trivial chiral flow is the matching of the Green-Schwarz couplings among the different theories. However, the pre-requisite understanding of Green-Schwarz anomaly cancellation in supercritical theories with tachyon condensation to chiral 10d string theories has not been worked out in the literature. Hence, we leave the computation of the relevant couplings, and their application to our junction configurations as an interesting direction for future research.

\section{Final remarks}
\label{sec:conclusions}

In this work we have initiated the exploration of explicit physical configurations of junctions of 10d string theories, which are predicted by the cobordism conjecture. Interestingly, we have shown that several non-trivial junction configurations can be described fairly explicitly using worldsheet techniques, which generalize in a natural way the notion of interpolation among 2d CFTs by going up and down the RG flow. Our generalization involves the appearance of non-compact CFTs at some point in the interpolation (as in \cite{Anastasi:2026cus}), and the appearance of a branch point, at which there are several mutually incompatible possibilities to continue the interpolation. We have provided explicit dynamical realizations of this picture, in which such parameters vary (and can appear and disappear) as we move along the spacetime branches of the junction. As in the case of 0A/0B and IIA/IIB domain walls in \cite{Anastasi:2026cus}, quantum corrections imply that the core of the junction is a strongly coupled lightlike defect. Hence, the ultimate fate of the junction lies beyond the reach of our worldsheet techniques; it is nevertheless satisfactory that our technniques describe the structure of the junction as precisely as possible for a worldsheet theory. In particular, it is amusing that the worldsheet theory predicts via quantum effects its own limit of applicability at the core region.

A simple application of our techniques, plus the addition of orbifold and orientifold quotients, leads to a rich set of junctions of 10d heterotic string theories, type 0 and type II theories, and quotients thereof. We have put special emphasis in the construction of bouquets, i.e. junctions of 10d string theories with chiral spacetime matter content, and non-trivial chiral flow across the junction, and we have explained their relation to the robustness of the junction against perturbations of the 2d theory by extra operators. We have found two remarkable bouquet configurations, in which the chiral fields of the theories in each branch flow across the junction into the others. These non-trivial chiral flows allow the configurations to admit 10d theories with non-anomalous chiral sets of fields defined on spacetimes bounded by the junction, without needing to resort to unknown symmetric mass generation mechanisms. Hence, one may envision that the topological protection of the junction, which cannot simply disappear without spoiling the chiral flow, may hold even in the UV completion of the strongly coupled core. Our results imply that the configuration is ready for such a UV complete transmissible junction to exist.

The chiral bouquet configurations open up other interesting new questions. One involves the junction of three 10d $E_8\times SO(16)$ heterotic theories. It would be interesting to understand the fate of the junction upon condensation of the 10d closed string tachyon. An even more impressive junction is the IIB bouquet, a 4-branch junction of 10d type II theory, type I, the $USp(32)$ theory and the $U(32)$ orientifold of 0B. These are the non-tachyonic descendants of 10d 0B theory under various orientifold and orbifold quotients. The intricate structure of the chiral flow in this junction suggests its protection may persist beyond the perturbative regime, so it may persist in other dual pictures. This would provide a fantastic tool to enrich the string duality relations of 10d supersymmetric strings to their non-supersymmetric cousins. In particular, the recent proposal for an M-theory description of 10d type 0A and 0B theories in \cite{Baykara:2026gem}, and their orientifolds \cite{Altavista:2026aar} may open up new ways to realize junctions of non-supersymmetric 10d string theories, possibly using analogues of the geometric constructions in \cite{Hellerman:2010dv}.

We have constructed our configurations using an underlying supercritical string interpretation for the extra degrees of freedom `up the RG flow'. However, we have emphasized that this supercritical interpretation is a mere (but very) convenient tool providing clean spacetime interpretations of the diverse operations, not an absolutely essential ingredient. Hence, even restricting to the class of cobordisms which can be described using worldsheet techniques, the class of junctions that can be build may be much larger than those constructed in this work. 

\begin{figure}[htb]
\begin{center}
\includegraphics[scale=.28]{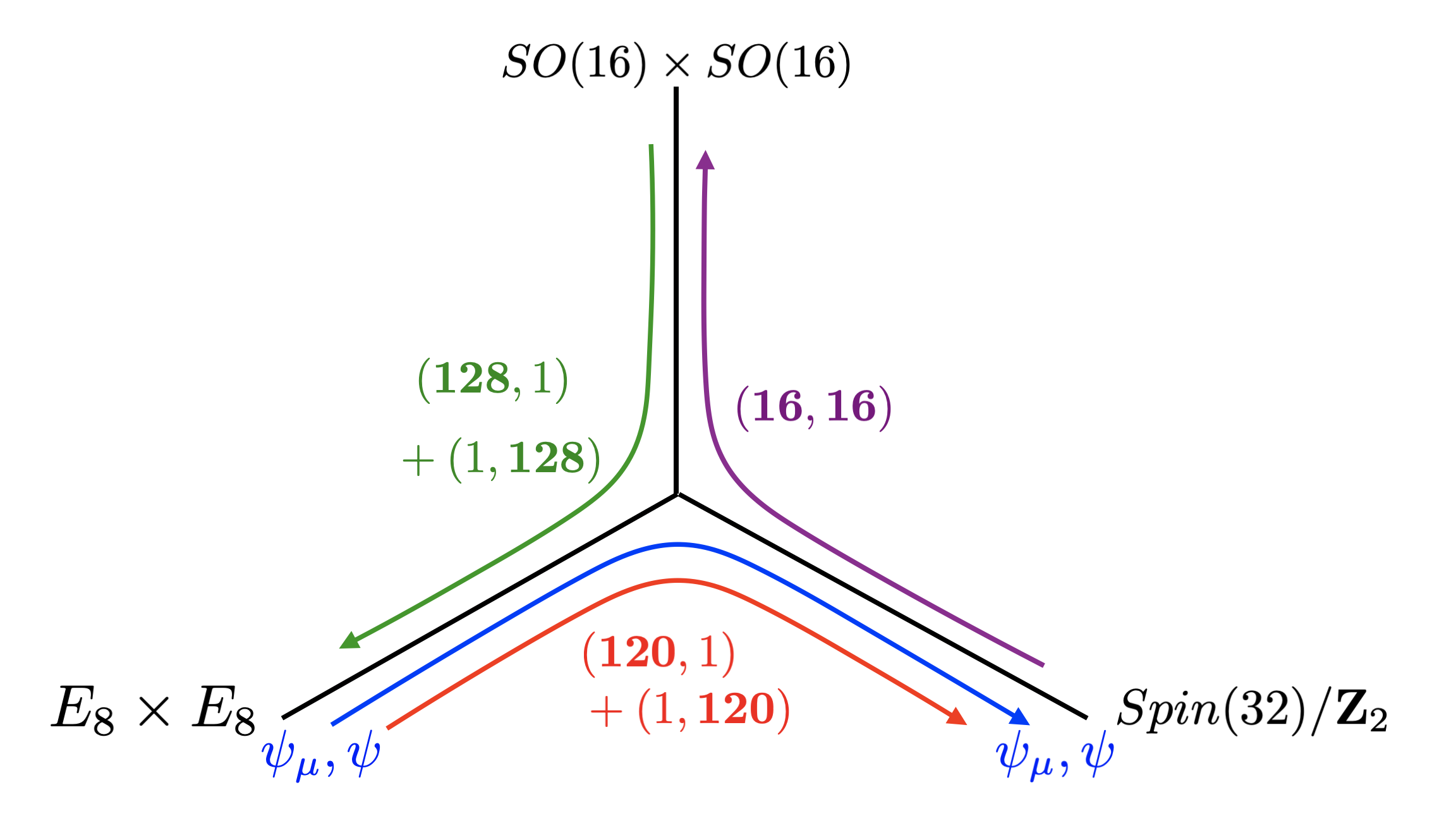}
\caption{\small The bouquet junction of the 10d supersymmetric $E_8\times E_8$ and the $Spin(32)/\IZ_2$ heterotic theories, and the 10d non-supersymmetric $SO(16)\times SO(16)$ theory. We display the chiral flow of gravitino/dilatinos (in blue), and chiral fermions in the different representations of the common subgroup $SO(16)\times SO(16)$ (in green, red and violet). Note that the orientations match to reproduce the familiar fact that the formal difference of the chiral spectrum of the two supersymmetric heterotic theories (i.e. the stacking of one with the chirality-flipped version of the other) corresponds to the chiral spectrum of the non-supersymmetric $SO(16)\times SO(16)$ theory.}
\label{fig:dream-bouquet}
\end{center}
\end{figure}

In particular, we expect the existence of a 3-branch bouquet junction between the 10d supersymmetric $E_8\times E_8$ and $Spin(32)/\IZ_2$ heterotic theories, and the 10d non-supersymmetric $SO(16)\times SO(16)$ theory \footnote{Shortly after the appearance of our work, a construction of this bouquet was provided by Tachikawa in \cite{Tachikawa:2026top}.}. The reason is that the familiar relations between their chiral spectra (once decomposed with respect to the common subgroup $SO(16)\times SO(16)$), amount to the  statement that they can be arranged as the chiral flow in a beautiful bouquet, as follows (see Figure \ref{fig:dream-bouquet}). First, the gravitino/dilatino of e.g. the $E_8\times E_8$ theory flow onto those of the $Spin(32)/\IZ_2$ theory; the gauginos of the $E_8\times E_8$ theory decompose as $({\bf 120},1)+(1,{\bf 120})$ and $({\bf 128},1)+(1,{\bf 128})$, while those of the $Spin(32)/\IZ_2$ decompose as $({\bf 120},1)+(1,{\bf 120})$ and $({\bf 16},{\bf 16})$. Hence the gauginos in the ${\bf 120}$'s of one theory can flow into those of the other; finally, the gauginos in the $({\bf 128},1)+(1,{\bf 128})$ of the $E_8\times E_8$ theory and those in the $({\bf 16},{\bf 16})$ of the $Spin(32)/\IZ_2$ theory can flow (with opposite spacetime chiralities) into the $SO(16)\times SO(16)$ theory branch, to fill out its 10d chiral fermion content. 

This junction would therefore provide a neat spacetime derivation of the relations between the chiral spectra of these theories (and hence of their anomaly polynomials, see e.g. \cite{Basile:2023knk}). Although this junction is not realized in the class of models in this work, the configuration is clearly topologically allowed, and in fact can be argued to admit a description within the context of TMFs \cite{Tachikawa:2024ucm}. A realization in the latter context would imply that such junction would be possible without involving strong coupling at the junction, consistently with the fact that the chiral flow in the bouquet guarantees no strongly coupled symmetric mass generation mechanisms are necessary.
It is tantalizing to envision that this may hold also for the other bouquet configurations we have encountered, given their consistent chiral flows.
 The use other (worldsheet or beyond) techniques to explore the realization of these bouquet configuration, and other similarly interesting ones, is a most promising direction for future work.

%%%%%%%%%%%%
%%%%%%%%%%%%

We envision the role of dynamical junctions of 10d string theories as building blocks of a zoo of string networks, of increasing complexity as new junctions are added by gluing, or of decreasing complexity as some external branches are glued together, naturally including ETW boundary configurations, and domain walls between pairs of theories. We hope the reader will join us in this exploration, where theories branch and intertwine into intricate bouquets of (chiral) flowers, still awaiting their ultimate arrangement.

%\newpage

\section*{Acknowledgments}

We are pleased to thank Jos\'e Calder\'on-Infante, Matilda Delgado, Miguel Montero, Salvatore Raucci, and Chuying Wang for useful discussions. This work is supported through the grants CEX2020-001007-S, PID2021-123017NB-I00 and  ATR2023-145703 funded by MCIN/AEI/10.13039/501100011033 and by ERDF A way of making Europe. The work by C. A. is supported by the fellowship LCF/BQ/DFI25/13000111 from ``La Caixa'' Foundation (ID 100010434). E. A. is supported by the fellowship LCF/BQ/DI24/12070005 from ``La Caixa'' Foundation (ID 100010434). R.A. acknowledges support from the ERC Starting Grant QGuide- 101042568- StG 2021, the Deutsche Forschungsgemeinschaft through the Collaborative Research Center 1624 “Higher Structures, Moduli Spaces and Integrability” and the Deutsche Forschungsgemeinschaft under Germany’s Excellence Strategy EXC 2121 Quantum Universe 390833306.
 
\newpage

\appendix

\section{Overview of 10d critical heterotic strings}
\label{app:critical}

In this appendix we quickly review the construction of critical 10d heterotic string theories, with special emphasis on the less familiar non-supersymmetric one. This is standard textbook background material (see e.g. \cite{Polchinski:1998rr}), but is included for completeness and as convenient preparation for the supercritical theories in appendix \ref{app:supercritical-heterotic}.

\subsection{The 10d diagonal non-supersymmetric $SO(32)$ heterotic}
\label{sec:10dheterotic-diagonal}

This is the simplest theory. We will work in the light cone gauge, in which the theory is described by 8 bosons $X^i$, their right-moving fermion partners ${\tilde \psi}^i$ and the 32 left-moving fermions $\lambda_a$. The theory is built by using the diagonal modular invariant partition function, namely the only $\IZ_2$ gauge symmetry is $(-1)^{F+{\tilde F}}$, i.e. overall fermion number so we sum over common spin structures for left and right fermions.

The spectrum contains the NSNS and the RR sector. The NSNS sector contains the massless graviton, 2-form, dilaton and $SO(32)$ gauge bosons, as well as a set of tachyons in the vector representation of $SO(32)$. The RR groundstate is massive, so there are no massless states in this sector.

\subsection{The 10d supersymmetric heterotic theories}
\label{sec:10d-susy-heterotics}

The two 10d supersymmetric heterotic strings are obtained by gauging further $\IZ_2$ symmetries, acting simultaneously on a multiple of 16 left-moving fermions, to overcome the $\IZ_{16}$-valued obstruction to attain modular invariant theories. 

Consider including the extra $\IZ_2$ gauging generated by $(-1)^{F}$, with $(-1)^{F}:\lambda_a\to -\lambda_a$, flipping the 32 left-moving fermions. A set of generators of the $\IZ_2\times\IZ_2$ symmetry are shown in table \ref{table:susy-so(32)}. Note that their product is $(-1)^{\tilde F}$, acting as a sum over spin structures for the right-movers and implementing the spacetime supersymmetric GSO projection. The resulting theory is the $Spin(32)/\IZ_2$ heterotic theory, which has independent sums over spin structures of the left- and of the right-movers. 

%%%%%%%%%%%%%%%%%%%%%%%%%%%%%%%%
\begin{table}[htb] \footnotesize
\renewcommand{\arraystretch}{1.25}
\begin{center}
\begin{tabular}{|c|c|c|}
\hline $SO(32)$   &  $\lambda_a$  & ${\tilde \psi}_i$ \\
\hline\hline 
$(-1)^{F+{\tilde F}}$  &  $-1$ & $-1$  \\
$(-1)^{F}$  &  $-1$ & $+1$  \\
%$(-1)^{\tilde F}$ & $+1$ & $-1$  \\
\hline \end{tabular}
\end{center} \caption{\small  A set of generators of the $\IZ_2\times \IZ_2$ gauge symmetry in the 10d supersymmetric $SO(32)$ heterotic.}
\label{table:susy-so(32)} 
\end{table}
%%%%%%%%%%%%%%%%%%%%%%%%%%%%%%%%%%%%

We can now further gauge a $\IZ_2$ symmetry acting as follows: split the 32 left-moving fermions in two sets of 16, denoted by $\lambda_A$, $\lambda'_{A'}$, with $A, A'=1,\ldots, 16$, and define $F_{16}$, $F_{16'}$ the fermion numbers for each set. We can quotient the above $Spin(32)/\IZ_2$ heterotic theory by the further $\IZ_2$ generated by $(-1)^{F_{16}}$. The action of a set of generators of the $\IZ_2\times\IZ_2\times\IZ_2$ symmetry are shown in table \ref{table:susy-e8e8}.

%%%%%%%%%%%%%%%%%%%%%%%%%%%%%%%%
\begin{table}[htb] \footnotesize
\renewcommand{\arraystretch}{1.25}
\begin{center}
\begin{tabular}{|c|c|c|c|}
\hline  $E_8\times E_8$  &  $\lambda_A$ &  $\lambda'_{A'}$ & ${\tilde \psi}_i$ \\
\hline\hline 
$(-1)^{F+{\tilde F}}$  &  $-1$ & $-1$ & $-1$  \\
$(-1)^{F}$  &  $-1$ & $-1$ & $+1$  \\
$(-1)^{F_{16}}$ & $-1$ & $+1$ & $+1$  \\
\hline \end{tabular}
\end{center} \caption{\small A set of generators of the $\IZ_2\times \IZ_2\times \IZ_2$ gauge symmetry in the 10d supersymmetric $E_8\times E_8$ heterotic.}
\label{table:susy-e8e8} 
\end{table}
%%%%%%%%%%%%%%%%%%%%%%%%%%%%%%%%%%%%

\subsection{The 10d non-supersymmetric heterotic strings}
\label{sec:10d-nonsusy-heterotics}

We now discuss several 10d non-supersymmetric heterotic string theories, with emphasis on the 10d $E_8\times SO(16)$ heterotic theory, which arises often in the constructions in the main text. Some original references are \cite{Dixon:1986iz,Seiberg:1986by,Alvarez-Gaume:1986ghj,Kawai:1986vd}, see \cite{Polchinski:1998rr} for a comprehensive treatment).

\subsubsection{The 10d non-supersymmetric $E_8\times SO(16)$ heterotic }
\label{sec:10d-nonsusy-e8so(16)}

Let us now go back to the original diagonal theory in section \ref{sec:10dheterotic-diagonal}, with the only $\IZ_2$ gauging associated to the overall left- plus right-moving fermion number. We now split the 32 fermions in two sets of 16, denoted by $\lambda_A$, $\lambda'_{A'}$, with $A, A'=1,\ldots, 16$ as above, and we gauge the $\IZ_2$ generated by $(-1)^{F_{16}}: \lambda_A \to -\lambda_A$ (with $\lambda'_{A'}$ invariant). The generators of the $\IZ_2\times\IZ_2$ gauge symmetry are shown in table \ref{table:susy-e8so(16)}. Note that the gauging of $(-1)^{F_{16}}$ implements a sum over spin structure of the $\lambda_A$, but the $\lambda'_{A'}$ and ${\tilde \psi}_i$ still share their spin structures (with sum implemented by $(-1)^{F+{\tilde F}+F_{16}}=(-1)^{F_{16'}+{\tilde F}}$.

%%%%%%%%%%%%%%%%%%%%%%%%%%%%%%%%
\begin{table}[htb] \footnotesize
\renewcommand{\arraystretch}{1.25}
\begin{center}
\begin{tabular}{|c|c|c|c|}
\hline  $E_8\times SO(16)$  &  $\lambda_A$ &  $\lambda'_{A'}$ & ${\tilde \psi}_i$ \\
\hline\hline 
$(-1)^{F+{\tilde F}}$  &  $-1$ & $-1$ & $-1$  \\
$(-1)^{F_{16}}$ & $-1$ & $+1$ & $+1$  \\
\hline \end{tabular}
\end{center} \caption{\small  A set of generators of the $\IZ_2\times \IZ_2$ gauge symmetry in the 10d non-supersymmetric $E_8\times SO(16)$ heterotic.}
\label{table:susy-e8so(16)} 
\end{table}
%%%%%%%%%%%%%%%%%%%%%%%%%%%%%%%%%%%%

Let us consider the light states. The different sectors (taking the NSNS sector as untwisted) and the corresponding periodicities are shown in Table \ref{table:bc-e8so16}. We discuss the untwisted and twisted sectors in turns.

%%%%%%%%%%%%%%%%%%%%%%%%%%%%%%%%
\begin{table}[htb] \footnotesize
\renewcommand{\arraystretch}{1.25}
\begin{center}
\begin{tabular}{|c|c||c|c|c||c|c|}
\hline 
Sector & Twist & $X_L^i$ & $\lambda_A$ & $\lambda'_{A'}$ & $X_R^i$ & ${\tilde \psi}^i$ \\
\hline 
(NS,NS;NS) & $1$ & $+1$ & $-1$ & $-1$ & $+1$ & $-1$ \\
(R,R;R) & $(-1)^{F+{\tilde F}}$ & $+1$ & $+1$ & $+1$ & $+1$ & $+1$\\
(R,NS;NS) & $(-1)^{F_{16}}$ & $+1$ & $+1$ & $-1$ & $+1$ & $-1$ \\
(NS,R;R) & $(-1)^{F+{\tilde F}}\cdot (-1)^{F_{16}}$ & $+1$ & $-1$ & $+1$ & $+1$ & $+1$\\
\hline
\end{tabular}
\end{center} \caption{\small Periodicities in the different sectors of the 10d $E_8\times SO(16)$ heterotic theory.
\label{table:bc-e8so16}}
\end{table}
%%%%%%%%%%%%%%%%%%%%%%%%%%%%%%%%%%%%

{\bf Untwisted sector:}\\
The RR sector contains only massive states, as in the parent diagonal theory, hence we skip its discussion. In the NSNS sector, the $\IZ_2= \langle (-1)^{F_{16}} \rangle$ projection leaves the massless graviton, 2-form, dilaton and projects the gauge bosons down to $SO(16)\times SO(16)$, and the tachyon to the vector representation of the $SO(16)$ associated to the $\lambda'_{A'}$ (i.e. the one not enhancing to $E_8$, see next paragraph). 

{\bf Twisted sector:}\\
In the sector with periodicities (R, NS; NS) for the fermions $\lambda_A$, $\lambda'_{A'}$, ${\tilde \psi}^i$. The groundstate is degenerate due to fermion zero modes for the $\lambda_A$, leading (after GSO projection) to a ${\bf 128}$ of the corresponding $SO(16)$. This is level-matched with a right-moving NS state in the vector of the spacetime $SO(8)$, resulting in 128 new gauge bosons which enhance the gauge group $SO(16)$ of the $\lambda_A$ to $E_8$. On the other hand, in the sector with periodicities (NS,R;R), the groundstate is degenerate due to fermion zero modes for the $\lambda'_{A'}$, leading to a ${\bf 128}+{\bf 128'}$ spinor of the corresponding $SO(16)$. This is level-matched with a right-moving R groundstate in the spinor ${\bf 8}_s+{\bf 8}_c$ of the spacetime $SO(8)$. The overall GSO projecion on these states fixes their overall chirality, producing states in the $({\bf 8}_s, {\bf 128})+({\bf 8_c},{\bf 128'})$ of the $SO(8)\times SO(16)$ Lorentz and gauge groups. 

Overall the spectrum has the structure
\vspace*{12pt}

\begin{tabular}{ccc}
$(\lambda_A,\lambda'_{A'};{\tilde \psi}^i)$  & State & Obs \\ \hline
(NS,NS;NS) & $\lambda_{-\frac 12}^{A'}|0\rangle \otimes |0\rangle$ &  Tachyon in  the $\fund$ of $SO(16)$\\ 
(NS,NS;NS) & $\alpha_{-1}^i|0\rangle \otimes \psi_{-\frac 12}^{j} |0\rangle$ & Graviton, 2-form, dilaton \\
(NS,NS;NS) & $\lambda_{-\frac 12}^A \lambda_{-\frac 12}^B |0\rangle \otimes \psi_{-\frac 12}^i|0\rangle$ & $SO(16)$ gauge bosons \\
(NS,NS;NS) & $\lambda'{}^{A'}_{-\frac 12} \lambda'{}^{B'}_{-\frac 12} |0\rangle \otimes \psi_{-\frac 12}^i|0\rangle$ & $SO(16)$ gauge bosons \\
(R,NS;NS) & $|{\bf 128}\rangle \otimes \psi_{-\frac 12}^i|0\rangle$ & Enhancement $SO(16)\to E_8$ \\
(NS,R;R)  & $|{\bf 128}\rangle \otimes |8_s\rangle+|{\bf 128'}\rangle \otimes |8_c\rangle$ & Chiral fermions in $({\bf 8}_s, {\bf 128})+({\bf 8_c},{\bf 128'})$\\ \hline
\end{tabular}

%\beqa
%(\lambda_A,\lambda'_{A'};{\tilde \psi}^i)\quad & {\rm State} & {\rm Obs.}\nonumber\\
%{\rm (NS,NS;NS)} & \lambda_{-\frac 12}^{A'}|0\rangle \otimes |0\rangle & {\rm Tachyon}\;{\rm in}\; {\rm the}\; \fund \; {\rm of}\; SO(16)\nonumber\\
%{\rm (NS,NS;NS)} &\alpha_{-1}^i|0\rangle \otimes \psi_{-\frac 12}^{j} |0\rangle & {\rm Graviton},\; {\rm 2-form},\; {\rm dilaton} \label{spectrum-diagonal}\\
%{\rm (NS,NS;NS)} &\lambda_{-\frac 12}^A \lambda_{-\frac 12}^B |0\rangle \otimes \psi_{-\frac 12}^i|0\rangle & SO(16) \; {\rm Gauge}\; {\rm boson} \nonumber\\
%{\rm (NS,NS;NS)} &\lambda'{}^{A'}_{-\frac 12} \lambda'{}^{B'}_{-\frac 12} |0\rangle \otimes \psi_{-\frac 12}^i|0\rangle & SO(16) \; {\rm Gauge}\; {\rm boson} \nonumber\\
%{\rm (R,NS;NS)} & |{\bf 128}\rangle \otimes \psi_{-\frac 12}^i|0\rangle & {\rm Enhanced}\; {\rm Group} \; SO(16)\to E_8 \nonumber\\
%{\rm (NS,R;R)}\quad  & |{\bf 128}\rangle \otimes |8_s\rangle+|{\bf 128'}\rangle \otimes |8_c\rangle & {\rm Chiral} \; {\rm Fermions} \; ({\bf 8}_s, {\bf 128})+({\bf 8_c},{\bf 128'})\nonumber
%\eeqa

\subsubsection{Other 10d non-supersymmetric heterotic theories}
\label{sec:10d-nonsusy-others}

For completeness we briefly discuss other 10d non-supersymmetric heterotic theories. Extending the idea used in the previous section \ref{sec:10d-nonsusy-e8so(16)}, one can continue including further $\IZ_2$ quotients generated by elements similar to $(-1)^{F_{16}}$, but acting on a different set of 16 fermions, provided that all the elements in the group also acts on 16-plets of fermions. A clever way of characterizing these possibilities is using the binary basis notation for the  labels of the 32 $\lambda_a$, in the form $d_1d_2d_3d_4d_5$, with $d_i=0,1$, namely $00000$, $00001$, $00010$,\ldots, $11111$ (adding a $+1$ is we wish to stick to the convention $a=1,\ldots, 32$ in decimal basis). We define $(-1)^{F_i}$ as the operator flipping the $\lambda_a$'s, with entry $d_i=0$, and take the group $(\IZ_2)^p$ generated by $p$ of these operators (with $p=1,2,3,4,5$), for which all elements act on some 16-plets of fermions. Starting from the original diagonal heterotic theory in section \ref{sec:10dheterotic-diagonal} and gauging the discrete group $(\IZ_2)^p$, we obtain a consistent heterotic theory for each value of $p$. The theory with $p=1$, with a single extra $\IZ_2$ gauging, is just the 10d non-supersymmetric $E_8\times SO(16)$ of section \ref{sec:10d-nonsusy-e8so(16)}. We refer to the others as the binary code 10d heterotic theories, and we skip their detailed discussion, directing the reader to \cite{Polchinski:1998rr} for details. 

Finally, there are variants of the theories in this and the previous sections which can be obtained by introducing discrete torsion. In particular, the 10d non-supersymmetric $SO(16)\times SO(16)$ heterotic theory is obtained by gauging the same discrete group used for the $E_8\times E_8$ heterotic in section \ref{sec:10d-susy-heterotics}, but including discrete torsion in the $\IZ_2\times \IZ_2$ generated by $(-1)^{\tilde F}$ and $(-1)^{F_{16}}$. This is equivalent to saying that the action of $(-1)^{F_{16}}$ on the sectors twisted by $(-1)^{\tilde F}$ has an additional $-1$ action, and viceversa. The discrete torsion implies that the states in NS and R sectors of right-movers now glue differently to left-moving states, resulting in a non-supersymmetric 10d theory. It is the only non-tachyonic non-supersymmetric 10d heterotic theory, but despite its interest, we skip its detailed discussion, and refer the reader to \cite{Polchinski:1998rr} for further details.

\section{Overview of supercritical type 0 and type II strings}
\label{app:supercritical}

In this Appendix we overview the construction of supercritical type 0 and type II string theories (for supercritical heterotic strings, see Appendix \ref{app:supercritical-heterotic}). These theories have a timelike linear dilaton, and so they always contain a strong coupling region in the far past (see \cite{Angius:2022mgh} for a proposed resolution in the supercritical bosonic theory). Otherwise, at late time they are weakly coupled and well behaved, and display interesting connections with the 10d critical string theories via closed tachyon condensation (see \cite{Hellerman:2006nx,Hellerman:2006hf,Hellerman:2006ff,Hellerman:2007fc,Hellerman:2007ym,Hellerman:2007zz,Hellerman:2008wp,Hellerman:2010dv,Berasaluce-Gonzalez:2013sna,Garcia-Etxebarria:2014txa,Garcia-Etxebarria:2015ota,Angius:2022mgh} for related work). In this quick review, we mainly follow \cite{Hellerman:2006ff}, to which we refer the reader for further details.

\subsection{Supercritical type 0 theories}
\label{app:type0}

The worldsheet theory of supercritical type 0 theory in flat $D=10+n$ dimensions are described (ignoring ghosts) by $10+n$ 2d real fields $(X^M, \psi^M)$ forming $(1,1)$ multiplets. We use untilded/tilded fields to denote the individual left- and right-moving fermions, if necessary. There is a timelike linear dilaton with gradient $V^M$, satisfying
\begin{equation}
V^M V_M=-\frac{n}{4\alpha'}.
\label{dilaton-back}
\end{equation} 
to ensure cancellation of the central charge. There is a gauge $\IZ_2$ symmetry generated by $(-1)^{F+{\tilde F}}$, where $F,{\tilde F}$ are the left- or right-moving worldsheet fermion number, leading to a diagonal modular invariant partition function. Equivalently, there is  sum  over spin structures common to both left- and right-moving sectors, which implements a GSO projection. We will focus on the case of even dimensions $n=2k$, for which there are two theories, the supercritical  0A and 0B theories, according to whether the GSO projections on the left- and right-moving Ramond groundstates are equal or opposite. The light spectrum in the NSNS sector contains a real closed tachyon scalar, and massless graviton, 2-form and dilaton fields, and the RR sector contains two sets of massless $p$-form potentials (with $p$ odd/even for the 0A/0B theories).

These theories provide an interesting arena to study closed tachyon condensation, as we now review. Following \cite{Hellerman:2006ff}, the tachyon couples as a worldsheet $(1,1)$ superpotential, 
\begin{equation}
\Delta {\cal L}\, =\, \frac{i}{2\pi}\int d\theta^+d\theta^-{\cal T}(X)\, =\; -\frac{1}{2\pi} \sqrt{\frac{\alpha'}{2}} F^M \partial_M {\cal T}(X) + 
\frac{i\alpha'}{4\pi} \partial_M\partial_N{\cal T}(X)\, \psi^M  {\tilde \psi}^N\, ,
\label{wsupo}
\end{equation}
where untilded and tilded fields correspond to left- and right-moving fields. Using the equations of motion for the auxiliary fields $F^M\sim G^{MN}\partial_N {\cal T}$, we get a worldsheet potential term
\begin{equation}
V\, =\, \frac{\alpha'}{16\pi} G^{MN}\partial_M{\cal T}\partial_N{\cal T}\, .
\label{pot}
\end{equation}
In \cite{Hellerman:2006ff} (see \cite{Hellerman:2006nx,Hellerman:2006hf,Hellerman:2006ff,Hellerman:2007fc,Hellerman:2007ym,Hellerman:2007zz,Hellerman:2008wp,Hellerman:2010dv,Berasaluce-Gonzalez:2013sna,Garcia-Etxebarria:2014txa,Garcia-Etxebarria:2015ota,Angius:2022mgh} for related work), a particular class of lightlike tachyon profiles was shown to lead to CFT's solvable in an $\alpha'$-exact way (1-loop exact in $\alpha'$, to be more precise), and to reduce the number of spacetime dimensions of the type 0 theory. 
For instance, we are interested in removing dimensions in pairs, by splitting the $n=2k$ extra dimensions in two sets $X^p$, $Y^p$, $p=1,\ldots, k$, and taking a tachyon profile 
\beqa
{\cal T}=\mu \exp(\beta X^+) \sum_p X^pY^p\, \, ,
\label{tachyon-quench1}
\eeqa
where for simplicity we have chosen the quadratic form to be diagonal, with $\mu$ being an arbitrary coefficient, and $\beta$ is fixed so that the tachyon deformation is an exactly marginal operator. In the following we often skip the exponential lightlike prefactor, but it should be implicitly kept in mind. Using (\ref{wsupo}) and (\ref{pot}), at late $X^+$ this tachyon profile corresponds to a mass term for the 2d boson $X$ and its fermion partner $\psi$, namely
\beqa
\Delta {\cal L}\, \sim \mu^2  (\sum_i [(X^p)^2+(Y^p)^2]+\mu \sum_i (\psi_X^p{\tilde\psi}_Y^p+\psi_Y^p{\tilde \psi}_X^p) \, .
\eeqa
Integrating out these massive 2d fields leads to the worldsheet content of the critical 10d theory. The quantum corrections induced by the massive modes renormalize the spacetime background and turn the originally timelike linear dilaton into a lightlike one, ensuring that the central charge continues to cancel  \cite{Hellerman:2006ff}. Although we will not need them, for completeness we mention that there exist more general transitions between supercritical theories in different dimensions; they are obtained by considering a lower rank $p<n$ quadratic form in (\ref{tachyon-quench1}), so that the endpoint of tachyon condensation is the $(D+2k-2p)$-dimensional theory, with a suitably renormalized timelike linear dilaton to continue ensuring cancellation of central charge.

\subsection{Supercritical theories related to 10d type II}
\label{app:typeII}

In this section we review two possible supercritical extensions of type II string theories, constructed in \cite{Hellerman:2004qa} and \cite{Hellerman:2004zm}. In either, the starting point is the supercritical type 0 theory, namely the light-cone gauge worldsheet content is given by $(8+n)$ bosons $X^{\hat i}$ and fermions $\psi^{\hat i}$, and a $\Z_2$ gauge symmetry of overall (left plus right) worldsheet fermion number. 

In order to construct a type II theory we need to gauge an additional $\Z_2$ acting as e.g. left-moving worldsheet fermion number. This extra $\Z_2$ symmetry is well known to have a $\IZ_8$ anomaly, so the supercritical versions of type IIA and type IIB  theories exist only for dimensions $D=2+8k$. Actually (see e.g. \cite{Hellerman:2004zm}) there are two different series of such theories, dubbed the standard series, for $D=10+16k$ (with partition function similar to that of 10d type II theories) and the Seiberg series, for $D=2+16k$ (which generalize the 2d type II theories in \cite{Seiberg:2005bx}). The mod-16 pattern in the structure of the partition function reflects the backfiring bosonization in \cite{BoyleSmith:2024qgx} (see also \cite{Heckman:2025wqd}), a subtle distinction between gauging the fermion parity symmetry (captured by a $\Z_{8}$ anomaly) and summing over spin structures (captured by a $\Z_{16}$ anomaly). These interesting theories are tachyon-free, but they do not arise in our constructions in the main text, so we skip their discussion and refer the reader to \cite{Hellerman:2004zm} for further details. 

We now turn to a second class of possible supercritical extensions of type II, which is valid for any even number of supercritical dimensions, $n=2k$ \cite{Hellerman:2004qa}, and which involve an orbifold acting on some of the supercritical bosons. Let us split the $(8+n)$ lightcone gauge coordinates into a set of $8$ denoted by $X^i$, and two sets of $k$, denoted by $X^p$, $Y^p$, $p=1,\ldots, n$. We similarly split the fermions into $8$ $\psi^i$, ${\tilde \psi^i}$, and two sets of $n$ denoted by $\chi_X^p$, ${\tilde \chi}_X^p$, and $\chi_Y^p$, $\tilde{\chi}_Y^p$, where untilded/tilded fields correspond to left-/right-movers. We consider the type 0 theory (with $\IZ_2$ gauge symmetry by overall fermion number, i.e. $(-1)^{F+{\tilde F}}$) and further orbifold the theory by $\IZ_2$ generated by $g_L=R (-1)^{F_{\psi}+F_{\chi_X}+{\tilde F}_{\chi_Y}}$, with a geometric action $R:Y^p\to -Y^p$ (leaving other bosons invariant), and hopefully obvious notation for the fermionic action. Note that the $g_L$ acts with a sign on the left-moving worldsheet supercurrent, so it defines an R-parity. In hindsight this will become the $\IZ_2$ quotient by left-moving fermion number upon the closed tachyon condensation discussed later. 

%%%%%%%%%%%%%%%%%%%%%%%%%%%%%%%%
\begin{table}[htb] \footnotesize
\renewcommand{\arraystretch}{1.25}
\begin{center}
\begin{tabular}{|c||c|c|c|c|c|c||c|c|c|c|c|c|}
\hline 
 Twist & $X_L^i$ & $X_L^p$ & $Y_L^p$ &  $\psi^i$ & $\chi_X^p$ & $\chi_Y^p$ & $X_R^i$  &  $X_R^p$ & $Y_R^p$ & ${\tilde \psi}^i$ & ${\tilde \chi}_X^p$ & ${\tilde \chi}_Y^p$ \\
\hline 
  $1$ & $+1$ & $+1$ & $+1$ & $-1$ & $-1$ & $-1$ & $+1$ & $+1$ & $+1$ & $-1$ & $-1$ & $-1$ \\
$(-1)^{F+{\tilde F}}$ & $+1$ & $+1$ & $+1$ & $+1$ & $+1$ & $+1$ & $+1$ & $+1$ & $+1$ & $+1$ & $+1$ & $+1$\\
$R (-1)^{F_{\psi}+F_{\chi_X}+{\tilde F}_{{\tilde\chi}_Y}}$ & $+1$ & $+1$ & $-1$ & $+1$ & $+1$ & $-1$ & $+1$ & $+1$ & $-1$ & $-1$ & $-1$ & $+1$\\
$R (-1)^{{\tilde F}_{\tilde \psi}+F_{\chi_Y}+{\tilde F}_{\tilde \chi_X}}$ & $+1$ & $+1$ & $-1$ & $-1$ & $-1$ & $+1$ & $+1$ & $+1$ & $-1$ & $+1$ & $+1$ & $-1$\\
\hline
\end{tabular}
\end{center} \caption{\small Periodicities in the different sectors of the orbifold of the supercritical type 0 theory (with the NSNS taken as trivially twisted).
\label{table:bc-orbifold0-supercritical}}
\end{table}
%%%%%%%%%%%%%%%%%%%%%%%%%%%%%%%%%%%%

Let us describe the light spectrum in the different sectors shown in Table \ref{table:bc-orbifold0-supercritical}. In the untwisted sector with respect to the new $\IZ_2$, we have the NSNS and RR fields of the underlying type 0 theory propagating in $D=10+2k$ dimensions, in combinations invariant under the orbifold action. Although all fields propagate in $(10+2k)$ dimensions, only those which remain dynamical locally at the  $(10+k)$-dimensional orbifold fixed locus will originate the spectrum of the 10d theory resulting under the tachyon condensation discussed below. These are the graviton, 2-form and dilaton, as well as a set of RR forms whose details depend on the spectific  theory (0A vs. 0B) and the number of extra dimensions (basically, odd/even degree forms for IIA/IIB theories).

In the sector twisted by $R (-1)^{F_{\psi}+F_{\chi_X}+{\tilde F}_{{\tilde\chi}_Y}}$, the states are localized at the orbifold fixed locus $Y^p=0$, which is $(10+k)$-dimensional. The left-moving groundstate is massless and degenerate, because of the fermion zero modes of $\psi^i$ and $\chi^p$, so it produces spinors of the $SO(8+k)$ local Lorentz group; it is level-matched with massless right-moving states obtained by applying a $-1/2$ modding oscillator of the orbifold odd fields (transforming in the vector representation of the local Lorentz $SO(8+k)$) to the tachyonic degenerate groundstate (producing spinors of the $SO(k)$ rotation group). The GSO projection due to overall fermion number correlates the overall chirality of the left- and right-moving spinors. The results is a $(10+k)$-dimensional set of states, transforming as vector representation of $SO(8+k)$ times a bi-spinor (of fixed overall chirality) of $SO(8+k)\times SO(k)$ (plus some extra states which disappear upon the eventual tachyon condensation, so we ignore them). In the sector twisted by $R (-1)^{{\tilde F}_{\tilde \psi}+F_{\chi_Y}+{\tilde F}_{\tilde \chi_X}}$ we get a similar structure up to exchange of left-/right-moving sectors, producing the same fields (with the opposite/same overall spinor chirality in the 0A/0B case). Upon tachyon condensation, these fields become the gravitinos/dilatinos of the resulting 10d Type II theories.

Let us now quickly discuss the tachyon condensation. The tachyon lives in the bulk spacetime, which is locally a type 0 theory, so it couples to the worldsheet as a $(1,1)$ superpotential, c.f. (\ref{wsupo}). The tachyon condensation down to 10d is triggered by the profile (\ref{tachyon-quench1}). We note that this interaction is invariant under the extra $\IZ_2$ symmetry, because  the superpotential is linear in the $\IZ_2$ odd fields $Y^p$, so it is odd, as should be for an R-parity (recall the $\IZ_2$ generator anticommutes with the left-moving supercharge).
This tachyon profile describes the removal of extra dimensions in pairs, and the 1-loop exact integration out of the massive degrees of freedom redefines the metric and dilaton background, so that the central charge readjusts, and we end up with a lightlike linear dilaton background of the critical 10d theory. Since the extra $\IZ_2$ gauging turns into $(-1)^{F}$, with $F$ the left-moving fermion number, we recover the supersymmetric GSO projection and the resulting theory is 10d Type II (A or B, according to whether the original $(10+2k)$-dimensional theory is the 0A or 0B theory). This construction will play an important role in the main text.

\section{Supercritical heterotic theories}
\label{app:supercritical-heterotic}

We now describe similar configurations for supercritical heterotic strings. There exist several supercritical versions of heterotic strings, considered e.g. in \cite{Hellerman:2004zm,Hellerman:2006ff}. They are all based on a $(0,1)$ worldsheet theory in $D=10+n$ flat dimensions with content (ignoring ghosts) given by $10+n$ bosons $X^M$, $10+n$ right-moving fermions ${\tilde \psi}^M$ and $32+n$ left-moving current algebra fermions $\lambda_{\hat a}$. There is a timelike linear dilaton to ensure cancellation of the central charge.

\subsection{The diagonal supercritical heterotic and closed tachyon condensation}
\label{sec:supercritical-heterotic-tachyons}

The simplest diagonal modular invariant theory, denoted as HO$^{+/}$ in \cite{Hellerman:2004zm,Hellerman:2006ff}, is based on the $\Z_2$ gauge symmetry generated by $(-1)^{F+{\tilde F}}$, where $F, {\tilde F}$ are the left- or right-moving worldsheet fermion numbers. The theory contains the graviton, 2-form and dilaton fields, as well as $SO(32+n)$ gauge bosons propagating in $10+n$ dimensions, and includes a set of tachyons ${\cal T}^{\hat a}$ in the vector representation of $SO(32+n)$. The Ramond groundstates are massive, so there are no massless fermions.

The theories in different dimensions are related by closed tachyon condensation. The spacetime tachyon profile ${\cal T}^{\hat a}(X)$ couples as a worldsheet $(0,1)$ superpotential
\beqa
\Delta {\cal L}=-\frac 1{2\pi}\int d\theta_+ \sum_{\hat a} \lambda_{\hat a} {\cal T}^{\hat a}(X)\, .
\label{supo-heterotic-tachyon}
\eeqa 
By expanding in components the right-moving superfields, and integrating out auxiliary fields, we obtain a scalar potential and fermion couplings
\beqa 
\Delta {\cal L}=-\frac 1{8\pi}\sum_{\hat a} ({\cal T}^{\hat a}(X))^2+\frac{i}{2\pi}\sqrt{\frac{\alpha'}2}\partial_M {\cal T}^{\hat a}(X)  \lambda_{\hat a}{\tilde\psi}^M\, .
\label{heterotic-tachyon-condensation}
\eeqa 
A particular class of lightlike closed tachyon condensation processes were shown in \cite{Hellerman:2006ff} to be tractable in an $\alpha'$-exact manner, leading to a reduction of the number of dimensions. Let us consider the particular case of removing $n$ supercritical dimensions to end up in a critical 10d theory. Splitting the $(8+n)$ bosons into 8 $X^i$ and $n$ $Y^p$, $p=1,\ldots,n$, and the right-moving fermions into 8 ${\tilde\psi}^i$ and $n$ ${\tilde\chi}_p$, and the $(32+n)$ left-moving fermions into 32 $\lambda_1$ and $n$ $\chi_p$, the tachyon profiles are of the form
\beqa
{\cal T}^{32+p}(X)=\mu\sqrt{\frac 2{\alpha'}}e^{\beta X^+} Y^{p}\, ,
\eeqa 
with $\beta$ again determined by the requirement that this is a marginal deformation. The above can be generalized to an arbitrary linear dependence of the tachyon with the coordinates, and for simplicity we have assumed it to be proportional to the identity, with coefficient $\mu$.

The resulting terms in (\ref{heterotic-tachyon-condensation}) become
\beqa 
\Delta{\cal L}=-\frac{\mu^2}{4\pi}e^{2\beta X^+} \sum_{p=1}^n \, (Y^{p})^2+\frac {i\mu}{2\pi}e^{\beta X^+} \sum_{p=1}^n [\, \chi_{p}\,(\,{\tilde\chi}_{p}+\beta Y^{p}{\tilde\psi}^+\,)\,]\, .
\label{interactions-heterotic-tachyon-condensation}
\eeqa 
At late $X^+$ the tachyon condensate removes the $n$ supercritical coordinates, as well as the extra fermions, hence the gauge symmetry becomes just $SO(32)$. Again, the one-loop corrections modify the metric and dilaton background such that the central charge adjusts to the new field content. In particular, the endpoint is a lightlike linear dilaton background of a critical 10d theory. Since the starting theory is the supercritical diagonal theory, the endpoint theory is the 10d non-supersymmetric $SO(32)$ diagonal theory, reviewed in section \ref{sec:10dheterotic-diagonal}. It is however easy to introduce extra gaugings to define supercritical theories which relate to other 10d supersymmetric heterotics via closed tachyon condensation, as we review in the next section.

\subsection{Supercritical heterotic with extra fermionic $\IZ_2$ gaugings}
\label{sec:supercritical-heterotic-orbifold-no}

It is easy to modify the above diagonal theory to produce new supercritical heterotic theories. The strategy is analogous to the construction of new 10d heterotic theories starting from the 10d diagonal one, c.f. Appendix \ref{app:critical}. Again the procedure is to gauge further $\IZ_2$ symmetries, acting simultaneously on sets of fermions with net chirality (left- minus right-) multiple of 16, to overcome the $\IZ_{16}$-valued constraint for modular invariance. Some such examples have been considered in \cite{Hellerman:2004zm,Hellerman:2006ff}, which fall in two classes, according to whether the $\IZ_2$ generators leave invariant or flip the bosons corresponding to the supercritical dimensions, equivalently, to whether they are Poincar\'e invariant in $D=10+n$ or just $10$ dimension (save for the timelike linear dilaton background). We review them in turns, and introduce some new variants necessary for the constructions in the main text. In this section we describe theories with the theories with no $\IZ_2$ orbifolds acting on the bosons, and postpone to section \ref{sec:supercritical-heterotic-orbifold-yes} the study of theories with non-trivial $\IZ_2$ action on bosons.

\subsubsection{A supercritical version of the 10d  supersymmetric $Spin(32)/\IZ_2$ theory}

We start with a class of theories, constructed in \cite{Hellerman:2004zm}, dubbed HO$^{+(n)}$ theories. It is obtained from the diagonal theory by splitting the $(32+n)$ left-moving fermions $\lambda_{\hat a}$ into $32$ $\lambda_a$ $a=1,\ldots,32$, and $n$ $\chi_p$, $p=1,\ldots,n$, respectively, and gauging by the $\IZ_2$ generated by $(-1)^{F_{32}}$ which flips $\lambda_a\to -\lambda_a$. 
The action of the generators is shown in table \ref{table:supercritical-ho+n}.

%%%%%%%%%%%%%%%%%%%%%%%%%%%%%%%%
\begin{table}[htb] \footnotesize
\renewcommand{\arraystretch}{1.25}
\begin{center}
\begin{tabular}{|c|c|c|c|c|}
\hline   HO$^{+(n)}$  &  $X^{\hat i}$ & $\lambda_a$ & $\chi_p$ & ${\tilde \psi}_{\hat i}$ \\
\hline\hline 
$(-1)^{F+{\tilde F}}$  & $+1$ &  $-1$ & $-1$ & $-1$  \\
$(-1)^{F_{32}}$ & $+1$ & $-1$ & $+1$ & $+1$  \\
\hline \end{tabular}
\end{center} \caption{\small  A set of generators of the $\IZ_2\times \IZ_2$ gauge symmetry in the supercritical HO$^{+(n)}$ heterotic.}
\label{table:supercritical-ho+n} 
\end{table}
%%%%%%%%%%%%%%%%%%%%%%%%%%%%%%%%%%%%

The splitting of the left-moving fermions, in two sets with different behavior under the $\IZ_2$'s, implies that the gauge group is reduced to $SO(32)\times SO(n)$.

We note that the action of the $\IZ_2$'s is compatible with the tachyon condensation process of the underlying theory in section \ref{sec:supercritical-heterotic-tachyons}. Tachyon condensation removes the $n$ extra supercritical bosons and right-moving fermions, and the $n$ left-moving fermion $\chi_p$. The action of the $\IZ_2$ gaugings on the worldsheet content of the remaining 10d theory is obtained from Table \ref{table:supercritical-ho+n} by erasing the action on the disappeared fields. It is easy to check that we recover the actions in Table \ref{table:susy-so(32)}, hence the endpoint of closed tachyon condensation is the 10d $Spin(32)/\IZ_2$ heterotic theory (as usual, with a lightlike linear dilaton background). The HO$^{+(n)}$ heterotic can thus be regarded as a supercritical version of the 10d supersymmetric $Spin(32)/\IZ_2$ heterotic theory.

\subsubsection{Supercritical version of general 10d heterotic theories}

It is now straightforward to generalize the procedure and build supercritical versions of other 10d heterotic theories. The procedure is simply to consider the worldsheet content and $\IZ_2$ gauge symmetries of the 10d heterotic theories in sections \ref{sec:10d-susy-heterotics} and \ref{sec:10d-nonsusy-heterotics}, and simply add $n$ extra bosons $X_p$ and right-moving fermions ${\tilde \psi}_p$, as well as $n$ extra left-moving fermions $\chi_p$, taking all of them to be invariant under all the $\IZ_2$ gauge symmetries (and including the timelike dilaton background to cancel their central charge). Because the extra field content is non-chiral under all the symmetries, the modification is automatically anomaly-free. Also, by construction, the theories admit a closed tachyon condensation in which the extra 2d fields are gapped and we return to the corresponding 10d theory. In other words, this procedure provides a simple supercritical version of the 10d $Spin(32)/\IZ_2$ and $E_8\times E_8$ theories, the $E_8\times SO(16)$ and binary code theories, and the $SO(16)\times SO(16)$ theory. These theories have a $(10+n)$-dimensional extension of the 10d spacetime fields, including the graviton, 2-form, dilaton, spacetime fermions (if present), and gauge bosons (with the same gauge symmetry as the 10d versions), and include the above mentioned tachyon triggering the return to 10d. The reduction of the $(10+n)$-dimensional spectrum down to 10d is difficult to track from the spacetime picture, given the failure of effective field theory to describe the process, but it is however straightforward from the worldsheet perspective.

\subsection{Theories with $\IZ_2$ actions on the supercritical bosons}
\label{sec:supercritical-heterotic-orbifold-yes}

We now consider another class of theories, obtained from the diagonal theory in section \ref{sec:supercritical-heterotic-tachyons} by gauging $\IZ_2$ symmetries which can act not only on the fermions, but also on the bosons. These theories are not Poincar\'e invariant in $D=(10+n)$ dimension, due to the orbifold action, and their spectrum splits into a bulk sector, and a twisted sector localized on the 10d orbifold fixed locus. We consider several possibilities in this class

\subsubsection{The $SO(32+n)$ theory}

We start with a class of theories, which for lack of a better name we dub the $SO(32+n)$ theory, constructed in \cite{Hellerman:2006ff} as follows. Let us first split the $(8+n)$ right-moving fermions into 8 fermions ${\tilde\psi}_i$, $i=1,\ldots, 8$ and $n$ fermions ${\tilde \chi}_p$, $p=1,\ldots, n$; we also split the $(8+n)$ bosons into 8 scalars $X^i$, $i=1,\ldots, 8$ and $n$ scalars $Y^p$, $p=1,\ldots, n$; finally, for later convenience we also split the $(32+n)$ left-moving fermions $\lambda_{\hat a}$ into 32 fermions $\lambda_a$, $a=1,\ldots, 32$ and $n$ fermions $\chi_p$, $p=1,\ldots, n$. 

We now gauge a $\IZ_2$ acting as $R(-1)^{F+{\tilde F}_{{\chi}_p}}$, where $R:Y^p\to -Y^p$, and $(-1)^F:\lambda_{\hat a}\to-\lambda_{\hat a}$, and $(-1)^{{\tilde F}_{{\chi}_p}}: {\tilde \chi}_p\to -{\tilde\chi}_p$ . Note that this action flips the $(32+n)$ left-moving fermions and $n$ right-moving fermions, so it acts chirally on 32 fermions, hence satisfying the $\IZ_{16}$ condition for modular invariance. The generators of the $\IZ_2\times\IZ_2$ discrete gauge symmetry are shown in table \ref{table:supercritical-so(32+n)}.

%%%%%%%%%%%%%%%%%%%%%%%%%%%%%%%%
\begin{table}[htb] \footnotesize
\renewcommand{\arraystretch}{1.25}
\begin{center}
\begin{tabular}{|c|c|c|c|c|c|c|}
\hline   $SO(32+n)$  &  $X^i$ & $Y^p$ & $\lambda_a$ & $\chi_p$ & ${\tilde \psi}_i$ & ${\tilde \chi}_p$ \\
\hline\hline 
$(-1)^{F+{\tilde F}}$  & $+1$ & $+1$ &  $-1$  & $-1$  & $-1$  & $-1$ \\
$R(-1)^{F+{\tilde F}_{{\chi}_p}}$ & $+1$ & $-1$ & $-1$  &  $-1$  & $+1$ & $-1$ \\
\hline \end{tabular}
\end{center} \caption{\small  A set of generators of the $\IZ_2\times \IZ_2$ gauge symmetry in the $SO(32+n)$ theory.}
\label{table:supercritical-so(32+n)} 
\end{table}
%%%%%%%%%%%%%%%%%%%%%%%%%%%%%%%%%%%%

We note that the product of the two generators in table \ref{table:supercritical-so(32+n)} acts as an R-parity on the right moving $(1,1)$ multiplets (i.e. bosons and their right-moving partners transform differently: the bosons $X^i$ are even, while ${\tilde \psi}_i$ are odd, and the $Y^p$ are odd, while the ${\tilde\chi}_p$ are even). This action is a supercritical version of $(-1)^{\tilde F}$ of the 10d critical theory in \ref{sec:10d-susy-heterotics}, as its action on the fermionic sector is by flipping just the 8 ${\tilde \psi}_i$.

Let us discuss the spectrum, in the untwisted and twisted sectors in turns.

%%%%%%%%%%%%%%%%%%%%%%%%%%%%%%%%
\begin{table}[htb] \footnotesize
\renewcommand{\arraystretch}{1.25}
\begin{center}
\begin{tabular}{|c||c|c|c|c||c|c|c|c|}
\hline 
 Twist & $X_L^i$ & $Y_L^p$ & $\lambda_a$ & $\chi_p$ &  $X_R^i$  & $Y_R^p$ & ${\tilde \psi}^i$ & ${\tilde \chi}^p$\\
\hline 
  $1$ & $+1$ & $+1$ & $-1$ & $-1$ & $+1$ & $+1$ & $-1$ & $-1$ \\
 $(-1)^{F+{\tilde F}}$ & $+1$ & $+1$ & $+1$ & $+1$ & $+1$ & $+1$ & $+1$ & $+1$\\
 $R(-1)^{F+{\tilde F}_{\tilde \chi}}$ & $+1$ & $-1$ & $+1$ & $+1$ & $+1$ & $-1$ & $-1$ & $+1$ \\
 $R(-1)^{{\tilde F}_{\tilde \psi^i}}$ & $+1$ & $-1$ & $-1$ & $-1$ & $+1$ & $-1$ & $+1$ & $-1$ \\
\hline
\end{tabular}
\end{center} \caption{\small Periodicities in the different sectors of the supercritical $SO(32+n)$ heterotic theory, with the NSNS periodicities taken as reference untwisted sector.}
\label{table:bc-so32+n-supercritical}
\end{table}
%%%%%%%%%%%%%%%%%%%%%%%%%%%%%%%%%%%%

The untwisted sector corresponds to the twists by $1$ and $(-1)^{F+{\tilde F}}$ in Table \ref{table:bc-so32+n-supercritical}, namely the NSNS and RR fields of the parent diagonal theory, which propagate in $D=(10+n)$ dimensions in a way invariant under the $\IZ_2$ geometric action $R$ flipping the extra coordinates $Y^p$. Note that even though the whole spectrum of the diagonal theory propagates in the bulk, only states even under $(-1)^{F+{\tilde F}_{\tilde \chi}}$ will be nonzero at the locus $Y^p=0$. In particular, only the graviton, 2-form, dilaton and gauge bosons of $SO(32)\subset SO(32+n)$ remain locally dynamical at the fixed locus (and the tachyon does not). These fields will hence survive the tachyon condensation process to 10d, described below.

The twisted sectors are localized at the 10d fixed locus $Y^p=0$, and morally correspond to NSR, RNS sectors (see Table \ref{table:bc-so32+n-supercritical} for the specific periodicities). In the sector twisted by $R(-1)^{F+{\tilde F}_{\tilde \chi}}$ all states are massive. In the sector twisted by $R(-1)^{{\tilde F}_\psi}$ the right-moving groundstate is massless and degenerate, due to fermion zero modes for the ${\tilde\psi}^i$, so it gives spinors ${\bf 8}_s+{\bf 8}_c$ of the Lorentz $SO(8)$; they can be level matched with a left-moving sector transforming as a vector ${\bf 8}_v$ of the spacetime $SO(8)$ or as current algebra states in the adjoint of the $SO(32)$. Including the GSO projection, we obtain localized 10d spacetime fermions transforming as vector-spinors of $SO(8)$, or as spinors in the adjoint of the $SO(32)$ locally unbroken at the orbifold fixed locus\footnote{There are in addition some additional fermions in the adjoint of $SO(n)$, which disappear in the tachyon condensation to 10d and will not be relevant for us.}. We dub these fields as gravitino/dilatinos and $SO(32)$ gauginos because they become those fields upon the tachyon condensation to 10d, to be discussed next.

This orbifold is compatible with the tachyon condensation in section \ref{sec:supercritical-heterotic-tachyons}, because the interactions (\ref{interactions-heterotic-tachyon-condensation}) are invariant. At late $X^+$, the tachyon profile removes the $n$ supercritical dimensions, so the endpoint is a critical 10d theory with a lightlike linear dilaton background. In the absence of the extra coordinates, the generator of the orbifold in the critical 10d slice is just $(-1)^{F}$, which flips the sign of the 32 left-moving current algebra fermions. Its product with $(-1)^{F+{\tilde F}}$ is  $(-1)^{\tilde F}$, with ${\tilde F}$ being the right-moving worldsheet fermion number, which implements the supersymmetric GSO projection. Hence, the resulting theory is the 10d supersymmetric $Spin(32)/\IZ_2$ heterotic theory (with a non-trivial lightlike linear dilaton background). Indeed, its spectrum arises from the supercritical perspective as follows. The untwisted spectrum contains the 10d graviton, 2-form and dilaton, and $SO(32)$ gauge bosons, while the twisted sector reduces to the 10d gravitino and dilatino, and the $SO(32)$ gauginos.

\subsubsection{Some variants of the $SO(32+n)$ theory by further $\IZ_2$ gaugings}

It is straightforward to consider the above $SO(32+n)$ theory and include further $\IZ_2$ gaugings acting only on sets of 16-plets of the left-moving fermions $\lambda_a$ to produce new consistent theories, which to our knowledge have not been considered in the literature. The theories are supercritical orbifold versions of the 10d heterotic strings built in sections \ref{sec:10d-susy-heterotics} and \ref{sec:10d-nonsusy-heterotics}. For instance, by taking the $SO(32+n)$ theory and gauging the $\IZ_2$ generated by $(-1)^{F_{32}}:\lambda_a\to -\lambda_a$, we obtain a supercritical orbifold theory with gauge group $SO(32)$, becoming the 10d supersymmetric $Spin(32)/\Z_2$ heterotic upon closed tachyon condensation. Further gauging the $\IZ_2$ generated by $(-1)^{F_{16}}$, flipping the sign of 16 out of these 32 fermions, we obtain a supercritical orbifold theory with gauge group $E_8\times E_8$, becoming the 10d supersymmetric $E_8\times E_8$ heterotic upon closed tachyon condensation (one can similarly introduce discrete torsion and obtain a supercritical orbifold version of the 10d $SO(16)\times SO(16)$ heterotic). Similarly, starting again with the $SO(32+n)$ theory it is possible to gauge the $\IZ_2$ generated by $(-1)^{F_{16}}$ only, and get a supercritical orbifold theory with gauge group $E_8\times SO(16+n)$, becoming the 10d non-supersymmetric $E_8\times SO(16)$ heterotic upon closed tachyon condensation. This latter strategy can be generalized to further $\IZ_2$ gaugings providing supercritical orbifold versions of the 10d binary code heterotic theories.

In short, these theories can be constructed by taking the worldsheet content and $\IZ_2$ gauge symmetries of a 10d heterotic theory, including the extra supercritical degrees of freedom $Y^p$, ${\tilde \chi}_p$, $\chi_p$, and further gauging by the $\IZ_2$ generated by the action $R(-1)^{F+{\tilde F}_\chi}$, where $R:Y^p\to -Y^p$, and $F+{\tilde F}_\chi$ flips the sign of the 32 $\lambda_a$, and the $n$ fermions $\chi_p$, ${\tilde\chi}_p$. The extra fields are invariant under the $\IZ_2$ gauge symmetries of the underlying 10d theory, and are non-chiral with respect to the new gauging, so the construction is consistent and modular invariant. By construction, the tachyon condensation process of the underlying $SO(32+n)$ theory survives to remove the extra fields and fall back to the 10d heterotic theory we started from. We skip further details of these constructions, and move a different kind.

\subsubsection{Symmetry breaking orbifold variants of the $SO(32+n)$ theory}
\label{sec:heterotic-orbifold-variants}

We now consider an alternative construction of variants of the $SO(32+n)$ theory, which to our knowledge have not appeared in the literature. The strategy is to take the fermionic action accompanying $R$ to leave invariant some subset of the $(32+n)$ left-moving fermions (in sets of 16-plets, to maintain modular invariance). If we take $16m$ fermions to be invariant, the gauge symmetry in the bulk is reduced from $SO(32+n)$ to $SO(32-16m+n)$. For large $n$ there may be many possible such theories, but we are interested in these theories to still admit the closed tachyon condensation to 10d. Since this process pairs up the left- and right-fermions $\chi_p$, ${\tilde\chi}_p$, they must transform in the same way under the $\IZ_2$, so the ${\tilde\chi}_p$ must flip sign under it (since the $\chi_p$ do). This implies that we are left with two theories, corresponding to $m=1,2$, which we dub the $SO(16)\times SO(16+n)$ and $SO(32)\times SO(n)$ variants, which we consider in turns.

\medskip

\subsubsection*{The $SO(16)\times SO(16+n)$ orbifold theory}

Let us now consider starting from the diagonal theory, and split the $(8+n)$ bosons into a set of 8 scalar $X^i$ and $n$ scalars $Y^p$, $p=1,\ldots, n$; we also split the $(8+n)$ right-moving fermions into 8 fermions ${\tilde \psi}^i$ and $n$ fermions ${\tilde \chi}_p$; we finally split the $(32+n)$ left-moving fermions into a set of 16+16 fermions $\lambda_A$, $\lambda'_{A'}$, $A, A'=1,\ldots, 16$, and $n$ fermions $\chi_p$. We now gauge the symmetry $R(-1)^{F_{\lambda'}+F_{\chi}+{\tilde F}_{\chi}}$. The generators are shown in table \ref{table:supercritical-so(16+n)}.

%%%%%%%%%%%%%%%%%%%%%%%%%%%%%%%%
\begin{table}[htb] \footnotesize
\renewcommand{\arraystretch}{1.25}
\begin{center}
\begin{tabular}{|c|c|c|c|c|c|c|c|}
\hline   $SO(16+n)$ variant  &  $X^i$ & $Y^p$ & $\lambda_A$ & $\lambda'_{A'}$ & $\chi_p$& ${\tilde \psi}_i$ & ${\tilde \psi}_p$ \\
\hline\hline 
$(-1)^{F+{\tilde F}}$  & $+1$ & $+1$ &  $-1$ & $-1$ & $-1$ & $-1$  & $-1$ \\
$R(-1)^{F_{\lambda'}+F_\chi+{\tilde F}_{\chi}}$ & $+1$ & $-1$  & $+1$ & $-1$  & $-1$  & $+1$ & $-1$ \\
\hline \end{tabular}
\end{center} \caption{\small  A set of generators of the $\IZ_2\times \IZ_2$ gauge symmetry  in the $SO(16+n)$ variant of the $SO(32+n)$ theory.}
\label{table:supercritical-so(16+n)} 
\end{table}
%%%%%%%%%%%%%%%%%%%%%%%%%%%%%%%%%%%%

%%%%%%%%%%%%%%%%%%%%%%%%%%%%%%%%
\begin{table}[htb] \footnotesize
\renewcommand{\arraystretch}{1.25}
\begin{center}
\begin{tabular}{|c||c|c|c|c|c||c|c|c|c|}
\hline 
 Twist & $X_L^i$ & $Y_L^p$ & $\lambda_A$ & $\lambda'_{A'}$ & $\chi_p$ &  $X_R^i$  & $Y_R^p$ & ${\tilde \psi}^i$ & ${\tilde \chi}^p$\\
\hline 
  $1$ & $+1$ & $+1$ & $-1$ & $-1$ & $-1$ &  $+1$ & $+1$ & $-1$ & $-1$ \\
 $(-1)^{F+{\tilde F}}$ & $+1$ & $+1$ & $+1$ & $+1$ & $+1$ & $+1$ & $+1$ & $+1$ & $+1$\\
 $R(-1)^{F_{\lambda'}+F_\chi+{\tilde F}_{\chi}}$ & $+1$ & $-1$ & $-1$ & $+1$ & $+1$ & $+1$ & $-1$ & $-1$ & $+1$ \\
 $R(-1)^{F_{\lambda}+{\tilde F}_{\psi}}$ & $+1$ & $-1$ & $+1$ & $-1$ & $-1$ & $+1$ & $-1$ & $+1$ & $-1$ \\
\hline
\end{tabular}
\end{center} \caption{\small Periodicities in the different sectors of the $SO(16+n)$ variant of the supercritical $SO(32+n)$ heterotic theory. 
\label{table:bc-so16+n-supercritical}}
\end{table}
%%%%%%%%%%%%%%%%%%%%%%%%%%%%%%%%%%%%

The set of elements in the orbifold group and the corresponding periodicities are shown in Table \ref{table:bc-so16+n-supercritical}. Note that upon tachyon condensation, the generators of the $\IZ_2\times\IZ_2$ become $(-1)^{F+{\tilde F}}$ and $(-1)^{F_{16}'}$, namely those of the 10d $ SO(16) \times E_8 $ heterotic theory.

Let us consider the untwisted sector with respect to the orbifold $\IZ_2$, namely the NSNS  sectors of the underlying diagonal theory (recall that RR states are massive). They lead to $(10+n)$-dimensional massless graviton, 2-form, dilaton and $SO(32+n)$ gauge bosons, and a tachyon in the vector representation of $SO(32+n)$, propagating in an orbifold invariant way. The local fields at the orbifold fixed locus are the massless graviton, 2-form, dilaton and $SO(16)\times SO(16+n)$ gauge bosons (hence the name), and the tachyon in the $\fund$ of $SO(16)$. 

Let us now consider the twisted sectors, lower two lines of Table \ref{table:bc-so16+n-supercritical}, which are localized in the 10d orbifold fixed locus $Y^p=0$, and hence transform under the 10d Poincar\'e group and the local $SO(16)\times SO(16+n)$ gauge symmetry. In the sector twisted by $R(-1)^{F_{\lambda'}+F_\chi+{\tilde F}_\chi}$ all states are massive. On the other hand, in the sector twisted by $R(-1)^{F_{\lambda}+{\tilde F}_{\psi^i}}$, the left-moving groundsates are massless and degenerate, giving spinors ${\bf 128}+{\bf 128'}$ of $SO(16)$. They are level-matched with the right-moving groundstate, which is also massless and degenerate, giving spinors ${\bf 8}_s+{\bf 8}_c$ of the Lorentz $SO(8)$. The GSO projection given by overall fermion number fixes the overall chirality of the surviving states, so we get states in the $({\bf 8}_s, {\bf 128})+({\bf 8_c},{\bf 128'})$ of the $SO(8)\times SO(16)$ Lorentz and gauge groups. Notice that this exactly the chiral content of the 10d $E_8\times SO(16)$ heterotic.

As explained, upon tachyon condensation the above theory turns into the 10d $E_8\times SO(16)$ theory. This raises the question of where the extra gauge bosons which enhance the symmetry to $E_8$ arise from, upon tachyon condensation. It is easy to see that they arise from the sector twisted by $R(-1)^{F_{\lambda'}+F_\chi+{\tilde F}_{\chi}}$, which only gives massive states before tachyon condensation, but produces massless states upon tachyon condensation. It is easy to check that even before the tachyon condensation they have the right structure: the left-moving sector (massive) groundstate transforms as an spinor of $SO(16+n)$ because of the fermion zero modes of $\lambda'_{A'}$, $\chi_p$, and these states can be level-matched with right-moving states obtained by taking the groundstate (which transforms as spinors of $SO(n)$) with one extra oscillator ${\tilde \psi}_{-\frac 12}^i$. These states correspond to massive vector fields transforming as spinors of $SO(16+n)$. Upon tachyon condensation, the $n$ extra units of rank disappear and the spinors of $SO(16)$ become massless, triggering the enhancement to $E_8$.

\medskip

\subsubsection*{The $SO(32)\times SO(n)$ orbifold theory}

We now consider a different variant, similar to the $SO(32+n)$ theory, but with the geometric action $R:Y^p\to-Y^p$ being accompanied by $(-1)^{F_\chi+{\tilde F}_\chi}$. Namely, the fermionic action flips only the supercritical fermions and leaves invariant the 32 left-moving fermions $\lambda_A$,$\lambda'_{A'}$. For simplicity, we gather the latter in 32 fermions $\lambda_a$, $a=1,\ldots, 32$. The generators are shown in table \ref{table:supercritical-so(n)}, and the periodicities in different sectors are in Table \ref{table:bc-son-supercritical}. We note that the new twist $R(-1)^{F_{\chi}+{\tilde F}_{\chi}}$ acts only on supercritical degrees of freedom, so upon tachyon condensation it becomes trivial and the theory becomes the 10d non-supersymmetric diagonal $SO(32)$ theory.

%%%%%%%%%%%%%%%%%%%%%%%%%%%%%%%%
\begin{table}[htb] \footnotesize
\renewcommand{\arraystretch}{1.25}
\begin{center}
\begin{tabular}{|c|c|c|c|c|c|c|}
\hline   $SO(n)$ variant  &  $X^i$ & $X^p$ & $\lambda_a$ & $\chi_p$ & ${\tilde \psi}_i$ & ${\tilde \psi}_p$ \\
\hline\hline 
$(-1)^{F+{\tilde F}}$  & $+1$ & $+1$ &  $-1$ & $-1$ & $-1$  & $-1$ \\
$R(-1)^{F_{\lambda_p}+{\tilde F}_{{\tilde \psi}_p}}$ & $+1$ & $-1$  & $+1$ & $-1$  & $+1$ & $-1$ \\
\hline \end{tabular}
\end{center} \caption{\small  A set of generators of the $\IZ_2\times \IZ_2$ gauge symmetry in the $SO(n)$ variant of the $SO(32+n)$ theory.}
\label{table:supercritical-so(n)} 
\end{table}
%%%%%%%%%%%%%%%%%%%%%%%%%%%%%%%%%%%%

%%%%%%%%%%%%%%%%%%%%%%%%%%%%%%%%
\begin{table}[htb] \footnotesize
\renewcommand{\arraystretch}{1.25}
\begin{center}
\begin{tabular}{|c||c|c|c|c||c|c|c|c|}
\hline 
 Twist & $X_L^i$ & $Y_L^p$ & $\lambda_a$ & $\chi_p$ &  $X_R^i$  & $Y_R^p$ & ${\tilde \psi}^i$ & ${\tilde \chi}^p$\\
\hline 
  $1$ & $+1$ & $+1$ & $-1$ & $-1$ & $+1$ & $+1$ & $-1$ & $-1$ \\
 $(-1)^{F+{\tilde F}}$ & $+1$ & $+1$ & $+1$ & $+1$ & $+1$ & $+1$ & $+1$ & $+1$\\
 $R(-1)^{F_{\chi}+{\tilde F}_{\tilde \chi}}$ & $+1$ & $-1$ & $-1$ & $+1$ & $+1$ & $-1$ & $-1$ & $+1$ \\
 $R(-1)^{F_{32}+{\tilde F}_{\tilde \psi^i}}$ & $+1$ & $-1$ & $+1$ & $-1$ & $+1$ & $-1$ & $+1$ & $-1$ \\
\hline
\end{tabular}
\end{center} \caption{\small Periodicities in the different sectors of the $SO(32)\times SO(n)$ variant of the supercritical $SO(32+n)$ heterotic theory. 
\label{table:bc-son-supercritical}}
\end{table}
%%%%%%%%%%%%%%%%%%%%%%%%%%%%%%%%%%%%

Let us discuss the spectrum. In the untwisted sector, RR states are massive, while in the NSNS sector we have the massless graviton, 2-form, dilaton and $SO(32+n)$ gauge bosons, and a tachyon in the fundamental of $SO(32+n)$, propagating in $D=10+n$ dimensions, in a way invariant under the orbifold action. The local symmetry at the orbifold fixed locus is $SO(32)\times SO(n)$, hence the name of this construction. The other dynamical fields at the fixed locus include the graviton, 2-form, dilaton, and a tachyon in the vector representation of $SO(32)$. 

In the sector twisted by  $R(-1)^{F_{\lambda}+{\tilde F}_{\tilde \psi}}$ all states are massive.
In the sector twisted by $R(-1)^{F_{\chi}+{\tilde F}_{\tilde \chi}}$, the left- and right-moving groundstates are tachyonic. The lightest level-matched states correspond to taking the right-moving groundstate, which is a spinor of $SO(n)$ times the left-moving states with one oscillator $\lambda_{-\frac 12}^a$ applied to the groundstate, which is a spinor of $SO(n)$. The GSO projection correlates the chiralities of the left- and right-moving $SO(n)$ spinors, so we obtain a set of 10d tachyons transforming in the vector representation of the local $SO(32)$, and as a bi-spinor of the local $SO(n)$. This becomes the tachyon of the 10d non-supersymmetric $SO(32)$ theory upon tachyon condensation.

\subsubsection*{Further variants}

It is straightforward to consider the $SO(32+n)$, the $SO(16)\times SO(16+n)$ or the $SO(32)\times SO(n)$ orbifold theories and implement further $\IZ_2$ gauging acting only on subsets of the 32 fermions $\lambda_A$, $\lambda'_{A'}$ (in a way compatible with the action of the existing orbifold) to produce new classes of supercritical orbifold variants of 10d heterotic theories. Since the new $\IZ_2$ actions leave the supercritical fields invariant, they are compatible with the closed tachyon condensation, which bring them back to the original 10d theories. In other words, such theories can be constructed by starting with the worldsheet content and $\IZ_2$ gauge symmetries of the 10d heterotic theories in sections \ref{sec:10d-susy-heterotics} and \ref{sec:10d-nonsusy-heterotics}, adding the supercritical set of $n$ bosons $Y^p$, and left- and right-moving fermions $\chi_p$, ${\tilde\chi}_p$, and orbifolding by $R(-1)^{f_m+F_\chi+{\tilde F}_\chi}$, where $(-1)^{f_m}$ is an action on the 32 fermions leaving $16m$ of them invariant, with $m=0,1,2$. We leave the discussion of the rich set of possibililites for the interested reader, and conclude here our study of old and new supercritical heterotic string theories.

\section{Details on the computation of the quantum correction}
\label{app:quantum}

In this section we provide some details of the computation of the 1-loop corrections to the effective action for the 2d boson $Z$ in the $Z$-branch, after integrating out the massive supermultiplets $X$ and $Y$. For simplicity we consider the superpotential in the absence of lightlike exponential in $X^+$, and refer the reader to \cite{Hellerman:2006ff} for details about its inclusion. The main conceptual modification, shown there, is that the result turns out to be 1-loop exact, so we restrict to this order.

Although the computation can be carried out in superspace, we do it in components. The starting point are the bosonic and fermionic lagrangian
\begin{equation}
\begin{split}
 L^B_{X,Y}&= \, \frac{1}{2} \eta^{\alpha \beta} \partial_{\alpha}X \partial_{\beta}X + \frac{1}{2}\eta^{\alpha \beta} \partial_{\alpha}Y \partial_{\beta}Y +\frac{1}{2} \mu(Z)^2(X^2+Y^2)\nonumber\\
    L^F_{X,Y}&= \, -i(\psi^{X} \partial_- \psi^{X} + \tilde{\psi^{X}}\partial_+ \tilde{\psi^{X}}) -i(\psi^{Y} \partial_- \psi^{Y} + \tilde{\psi^{Y}}\partial_+ \tilde{\psi^{Y}}) + \mu (\tilde{\psi^{X}} \psi^{Y} + \tilde{\psi^{Y}} \psi^{X})\, , \\
\end{split}
 \end{equation}
where $\mu(Z)$ is the $Z$-dependent mass for bosons and fermions. Since the massive spectrum is fully supersymmetric, fermion-boson degeneracy implies cancellation of corrections to the scalar potential for $Z$. In the path integral formulation, this follows from the equality of the bosonic and fermionic determinants computing the 1-loop vacuum amplitude.

In order to compute the corrections to the kinetic term, we expand the field $Z$ into background and fluctuations, 
\beqa
Z(\sigma) &= Z_0 + \delta Z(\sigma) \, .
\eeqa
We introduce the notation for the expansion of the $Z$-dependent mass terms
\begin{equation}
\begin{split}
\mu^2(Z) &= \,  \mu_0^2 + m_1 \,\delta Z(\sigma) + \frac{1}{2}m_2(\delta Z(\sigma))^2 \nonumber \\
\mu(Z)&= \,\mu_0+ \mu_1\,\delta Z(\sigma)\, ,  \\
\end{split}
\end{equation}
with the relations between coefficients spelled out later on.

The relevant diagrams are loops of the massive bosons and fermions, with two external legs of $\delta Z$ carrying momentum $k$. The bosonic diagram is given by
\begin{equation}
      -\frac{1}{2} m_1^2 \delta Z(k) \delta Z (-k) \Pi_b(k)\quad,\; {\rm with} \, \, \,  \, 
     \Pi_b(k)= \int \frac{d^2 p}{(2\pi)^2}  \frac{1}{(p^2 + \mu_0^2)((k+p)^2 + \mu_0^2)}\, .
\end{equation}
Computing the diagram using Feynman parametrization etc, one obtains a correction to the metric given by 
\begin{equation}
    \frac{m_1^2}{24 \pi \mu_0^4}\, .
\end{equation}
The fermionic lagrangian has a mixed mass term, but it can be diagonalized into two combinations with mass eigenvalues $\pm \mu_0$. For corresponding fermionic loop diagram for $+\mu_0$ combination is
\begin{equation}
\begin{aligned}
&\frac{\mu_1^2}{4}\,\delta Z(k)\delta Z(-k)\Pi_f(k)\quad,\; {\rm with} \; 
     \Pi_f(k)=\int\frac{d^2p}{(2\pi)^2}\frac{\operatorname{Tr}_{\alpha\beta}\!\Big[ (\mu_0 - i \slashed{p})(\mu_0 -i(\slashed{p} + \slashed{k}))\Big]}{(p^2 + \mu_0^2)((p+k)^2 + \mu_0^2)}\, .
\end{aligned}
\end{equation}
Using the Euclidean Clifford algebra in 2d, we have 
\begin{equation}\label{eq: traccia}
 \operatorname{Tr}_{\alpha\beta}\!\Big[ (\mu_0 - i \slashed{p})(\mu_0 -i(\slashed{p} + \slashed{k}))\Big]=2(\mu_0^2-p(p+k)) \, .  
\end{equation}
Since we are only left with terms quadratic in $\mu_0$, the contribution from the fermion of mass $-\mu_0$ is identical, leading to an extra factor of 2. Evaluating the integral using Feynman parametrization etc, one obtains a correction to the metric from the two fermions given by
\begin{equation} \label{eq: correction fermionic metric}
    \frac{\mu_1^2}{12 \pi \mu_0^2}
\end{equation}
Using the relations
\beqa
m_1=  \mu_0 \mu_1 \quad , \quad m_2= 2\mu_1^2\, ,
\eeqa
the full correction to the kinetic term is
\begin{equation}
  \frac{\mu_1^2}{6\pi \mu_0^2}+\frac{\mu_1^2}{12\pi \mu_0^2}=\frac{\mu_1^2}{4\pi \mu_0^2}\, ,
\end{equation}
which reproduces the $(\partial Z)^2/Z^2$ terms leading to infinite distance to the origin, as explained in the main text. The similar computations when the field-dependent mass term includes the exponential lightlike dependence in $X^+$ can be worked out similarly, we refer the reader to \cite{Hellerman:2006ff} for further details.

\section{Junction conditions from correlators: A case example}
\label{app:correlators}

In this Appendix we provide some details on the selection rules for the correlators encoding the junction conditions for the 3-branch bouquet of three 10d $E_8\times SO(16)$ heterotic theories in section \ref{sec:junction-conditions-heterotic}.

Recall from section \ref{sec:3-branch-e8-so16} the split of the 34 left-moving fermions into 8 $\lambda_{A_1}$, 8 $\lambda_{A_2}$, 8 $\lambda'_{A'_1}$, 7 $\lambda'_{A'_2}$, and $\lambda_X$, $\lambda_Y$, $\lambda_Z$. Let us denote by $SO(8)_X$, $SO(8)_Y$, $SO(8)_Z$ and $SO(7)$ the groups associated to $\lambda_{A_1}$, $\lambda_{A_2}$, $\lambda'_{A'_1}$, and $\lambda'_{A'_2}$, respectively. The notation is motivated because on the $X$-branch, the 10d $SO(16)$ gauge group (which we denote by $SO(16)_X$) arises from the enhancement of $SO(8)_X\times SO(7)$ together with $\lambda_X$, and similarly for the other branches. Complementarily, the 10d $E_8$ on the $X$-branch (which we denote by $(E_8)_X$) arises from the $SO(8)_Y\times SO(8)_Z$ (together with $\lambda_Y$, $\lambda_Z$, which are gapped upon tachyon condensation), and similarly for the other branches. Let us similarly recall that the 10 right-moving fermions are split into 7 ${\tilde \psi}^i$, and ${\tilde\psi}_X$, ${\tilde\psi}_Y$,${\tilde\psi}_Z$, and recall that on the $X$-branch, the 10d spacetime (or rather its light-cone gauge $SO(8)$, which we denote by $SO(8)'_X$) is associated to the enhancement of the $SO(7)'$ of the  ${\tilde \psi}^i$ with ${\tilde\psi}_X$, and similarly for the other branches.

Let us introduce a vector notation\footnote{Replacing the  representations of non-abelian groups by their weights, this is largely analogous to the momentum lattice in the bosonized version of this fermion system, except for spelling out the structure of fermion zero modes for unpaired fermions.} indicating the quantum numbers under the left-moving $SO(8)_X$, $SO(8)_Y$, $SO(8)_Z$, $SO(7)$, and the occupation numbers of $\lambda_X$, $\lambda_Y$, $\lambda_Z$ and the right-moving $SO(7)'$, and occupation numbers of ${\tilde\psi}_X$, ${\tilde\psi}_Y$,${\tilde\psi}_Z$. For instance, a chiral fermions in the $X$-branch in the $({\bf 128},{\bf 8}_s)$ of the $SO(16)_X\times SO(8)'_X$ gauge and Lorentz groups corresponds to
\beqa
&& ({\bf 8}_{s,X}, {\bf 1}, {\bf 1}, {\bf 8})\oplus (\textstyle{+\frac 12},0,0)\oplus {\bf 8}\oplus(\textstyle{+\frac 12},0,0)\nonumber \\
&& ({\bf 8}_{c,X}, {\bf 1}, {\bf 1}, {\bf 8})\oplus (\textstyle{-\frac 12},0,0)\oplus {\bf 8}\oplus (\textstyle{-\frac 12},0,0)\, .
\label{momentum-spinor}
\eeqa
Here the first four entries indicate the representations under the $SO(8)^3\times SO(7)$ left-moving current algebra; the next three entries indicate the zero mode structure of
$\lambda_X$, $\lambda_Y$, $\lambda_Z$ ($\pm 1/2$ indicate the fermion zero mode is in the vertex operator or not, and $0$ implies there exists no fermion zero mode in that sector); similarly, the following entry is the Lorentz $SO(7)'$ quantum number, and the final three entries indicate the zero mode structure of ${\tilde\psi}_X$, ${\tilde\psi}_Y$,${\tilde\psi}_Z$ (with the same convention as for the left-moving sector).

Similarly, a chiral fermion in the $X$-branch in the $({\bf 128}',{\bf 8}_c)$ of the $SO(16)_X\times SO(8)'_X$ gauge and Lorentz groups is encoded in
\beqa
&& ({\bf 8}_{s,X}, {\bf 1}, {\bf 1}, {\bf 8})\oplus (\textstyle{-\frac 12},0,0)\oplus {\bf 8}\oplus (\textstyle{-\frac 12},0,0)\nonumber \\
&& ({\bf 8}_{c,X}, {\bf 1}, {\bf 1}, {\bf 8})\oplus (\textstyle{+\frac 12},0,0)\oplus {\bf 8}\oplus(\textstyle{+\frac 12},0,0)\, .
\label{momentum-spinor-conjugate}
\eeqa
Namely it is related to (\ref{momentum-spinor}) by a chirality flip for left- and right-moving sectors (implemented by adding/removing one fermion zero mode $\lambda_X$, ${\tilde\psi}_X$).

Finally, another state of interest in the $X$-branch is the 10d $(E_8)_X$ gauge boson in the spinor of the underlying $SO(16)$ before the enhancement. Recall from section \ref{sec:heterotic-orbifold-variants} that in the supercritical theory this arises from a gauge boson in the spinor of the supercritical $SO(16+n)$ which becomes massless upon tachyon condensation. This state corresponds to
\beqa
&& ({\bf 1}, {\bf 8}_{s,Y}, {\bf 8}_{s,Z}, {\bf 1}) \oplus \, \pm (0,\textstyle{+\frac 12},\textstyle{+\frac 12}) \oplus ({\bf 7}+1) \oplus\, \pm (0,\textstyle{+\frac 12}, \textstyle{+\frac 12})\nonumber \\
&& ({\bf 1}, {\bf 8}_{c,Y}, {\bf 8}_{c,Z}, {\bf 1}) \oplus \, \pm (0,\textstyle{+\frac 12},\textstyle{+\frac 12}) \oplus ({\bf 7}+1)\oplus \,\pm (0,\textstyle{+\frac 12}, \textstyle{+\frac 12})\nonumber \\
&& ({\bf 1}, {\bf 8}_{s,Y}, {\bf 8}_{c,Z}, {\bf 1}) \oplus \, \pm (0,\textstyle{+\frac 12}, \textstyle{-\frac 12}) \oplus ({\bf 7}+1)\oplus \,\pm (0,\textstyle{+\frac 12}, \textstyle{-\frac 12})\nonumber \\
&& ({\bf 1}, {\bf 8}_{c,Y}, {\bf 8}_{s,Z}, {\bf 1}) \oplus \, \pm (0,\textstyle{+\frac 12}, \textstyle{-\frac 12}) \oplus ({\bf 7}+1)\oplus \,\pm (0,\textstyle{+\frac 12}, \textstyle{-\frac 12})\, .
\eeqa
We are now ready to check the selection rules of 3-point functions. The a priori large scan of possibilities simplifies by noting that chiral fermions arise in twisted sectors. Then the orbifold selection rules imply that two states in $g_{16,X}$ and $g_{16,Y}$ twisted sectors must interact with a third in the sector twisted by $g_{16,Z}$ (as we see below, this is simply encoded in sum rules in our vector notation). This explains why it makes sense to consider the correlator between chiral spinors $({\bf 128},{\bf 8}_s)_X$, $({\bf 128},{\bf 8}_s)_Y$ and an $(E_8)_Z$ gauge boson in the above kind of spinor representation, which arise in the corresponding twisted sectors, and propagate on the corresponding branches, as indicated by their subindex. Indeed, we can provide an illustrative example of a non-zero correlator involving e.g. the states 
\beqa
&& ({\bf 128},{\bf 8}_s)_X \supset ({\bf 8}_{s,X}, {\bf 1}, {\bf 1}, {\bf 8})\oplus (\textstyle{-\frac 12},0,0)\oplus {\bf 8}\oplus (\textstyle{-\frac 12},0,0)\nonumber \\
&& ({\bf 128},{\bf 8}_s)_Y \supset ( {\bf 1}, {\bf 8}_{s,Y},{\bf 1}, {\bf 8})\oplus (\textstyle{0,-\frac 12},0)\oplus {\bf 8}\oplus (\textstyle{0,-\frac 12},0) \\
&& ({\bf 128},{\bf 8}_v)_Z \rightarrow ({\bf 8}_{s,X}, {\bf 8}_{s,Y}, {\bf 1},  {\bf 1}) \oplus \,  (\textstyle{-\frac 12},\textstyle{-\frac 12},0) \oplus ({\bf 7}+1) \oplus\,  (\textstyle{-\frac 12}, \textstyle{-\frac 12},0)\, .\nonumber
\eeqa
In other words, the quantum numbers of the combination of the first two and the third, by simply contracting the left-moving $SO(7)'$ spinor to a singlet, and the right-moving $SO(7)$ spinor to the vector (plus a singlet, filling out an ${\bf 8}_v$ of the Lorentz $SO(8)'_Z$). A similar discussion can be carried out for other components in the corresponding representations. Finally, by the permutation symmetry of the branches, we have a similar pattern for other combinations of spinors in any two other branches and a gauge boson in the third. 

Clearly, a similar analysis can be carried out for the fermions in the $({\bf 128'},{\bf 8}_c)$ in a given branch, which turn into fermions in the $({\bf 128'},{\bf 8}_c)$ in a second branch, by emitting an $E_8$ gauge boson in the third branch. We also note that it is not possible for a $({\bf 128},{\bf 8}_s)$ in one branch to turn into a $({\bf 128'},{\bf 8}_c)$ in another, or viceversa, since there is no $E_8$ gauge boson with the correct quantum numbers in the third.

As explained above, the correlator in the supercritical theory descends to a correlator in the junction of 10d strings upon tachyon condensation (which merely amounts to deleting information from the corresponding supercritical directions in each branch). The conclusion is that, as anticipated, a chiral fermion of the 10d $E_8\times SO(16)$ theory in a given branch can continue propagating across the junction as a chiral fermion of the 10d $E_8\times SO(16)$ theory in a second branch, with the simultaneous emission of an $E_8$ gauge boson (in a suitable spinor representation of the underlying $SO(16)$) of the 10d $E_8\times SO(16)$ theory in the third branch.

\bibliographystyle{JHEP}
\bibliography{refs}

\end{document}